\documentclass[preprint,aps,prc,showpacs]{revtex4}

\usepackage{graphicx}
\usepackage{bm}

\begin{document}

\title{ Kaon production and propagation at intermediate relativistic 
        energies%
        \footnote{Work supported by GSI Darmstadt} }
\author{A.B. Larionov$^{1,2}$ and U. Mosel$^1$}

\affiliation{$^1$Institut f\"ur Theoretische Physik, Universit\"at Giessen,
              D-35392 Giessen, Germany\\
$^2$RRC "I.V. Kurchatov Institute", 123182 Moscow, Russia}

\date{\today}

\begin{abstract}
We systematically study $K^+$ observables in nucleus-nucleus collisions at 
1-2 A GeV within the Boltzmann-Uehling-Uhlenbeck (BUU) transport model. 
We compare our calculations with the KaoS data on the kaon multiplicities 
and spectra. In addition, the kaon collective flow is computed and
compared with the FOPI and KaoS data.
We show, that the elliptic kaon flow measured recently by the
KaoS Collaboration is best described by using the Brown-Rho 
parametrization of the kaon potential ($U_K(\rho_0) \simeq 30$ MeV).  
\end{abstract}


\maketitle

\section{ Introduction }

Since about 20 years strangeness production in heavy-ion collisions is
a hot topic of theoretical and experimental studies. Due to rather
high energy thresholds in NN collisions
($E_{beam}=1.58$ GeV for $N N \to K \Lambda N$ and $E_{beam}=2.5$ GeV
for $N N \to N N K \bar K$) the secondary processes $\Delta N \to K Y N$,
$\pi N \to K Y$ and $\pi Y \to \bar K N$ -- which require high baryon
density -- are important in the case of nucleus-nucleus collisions 
at 1-2 A GeV.
Moreover, due to the relatively low ($\sim 10$ mb) $KN$ scattering cross
section and the absence of the absorption channel of a kaon on a nucleon
in strong interactions, nuclear matter is practically transparent for
kaons \cite{RK80}. Thus, the kaon yield is a good probe for the nuclear 
equation-of-state (EOS). This idea has originally been proposed and tested 
in BUU calculations by Aichelin and Ko in Ref. \cite{AK85}. Recently it has 
been shown by Fuchs et al. \cite{Fuchs01} within the T\"ubingen QMD model, 
that the ratio of the kaon yields in Au+Au and C+C collisions plotted vs the 
beam energy favours a soft EOS (K=200 MeV) when comparing with the KaoS data 
\cite{Sturm01}. This ratio seems to be weakly sensitive to the experimentally 
not measurable $\Delta N \to K Y N$ cross section and to the choice of
the kaon potential and serves as a good probe for the nuclear EOS.

Another direction of studies is related to the kaon and antikaon propagation
in the nuclear medium 
\cite{Li95,Li951,Li96,Li961,Brat97,David99,Mish04,Zheng04}.
For these studies the $K$ and $\bar K$ mean field potentials play a
crucial role. In the lowest order approximation to the chiral Lagrangian
the kaon (antikaon) potential has an attractive scalar and a repulsive
(attractive) vector part \cite{KN86,NK87,Li951}. This leads to a weakly 
repulsive potential ($U_K(\rho_0) \simeq 7$ MeV, where $\rho_0=0.17$ fm$^{-3}$)
for kaons and a strongly attractive potential ($U_{\bar K}(\rho_0) \simeq -100$
MeV) for antikaons. The kaon \cite{Li95,Li951} and the antikaon 
\cite{Li961} in-plane flows and the kaon azimuthal distributions \cite{Li96}
are strongly influenced by the $K$ and $\bar K$ potentials.
The authors of Ref. \cite{Brat97} have shown within the HSD model, that a
repulsive potential ($U_K(\rho_0) \simeq 30$ MeV) seems to be needed for 
description of the first FOPI data \cite{Herr96} on the $K^+$ in-plane flow.
In the most recent HSD analysis of Ref. \cite{Mish04}, where different
kaon potentials were tested, the one given by the chiral perturbation
theory in the relativistic Hartree approximation ($U_K(\rho_0) \simeq 20$
MeV) was found to give the best agreement with the kaon flow data.
The same value of the kaon potential at normal nuclear matter density 
($U_K(\rho_0) = 20\pm5$ MeV) was reported in a recent CBUU analysis of the 
$K^+$ production in proton-nucleus reactions by Rudy et al. \cite{Rudy05}.

In the approaches of Refs. \cite{Li95,Li951,Li96,Brat97,Mish04,Rudy05} the 
spatial components of the vector field in the kaon potential were 
neglected which should be approximately valid in a central zone of a symmetric
colliding system where the baryon current disappears. 
The spatial components were taken into account by Fuchs et al. \cite{Fuchs98}
within the T\"ubingen QMD model  
and led to a much smaller negative flow of $K^+$'s. This is a consequence of 
the cancellation effect of the repulsive time component of the kaon vector 
field by the Lorentz force generated by the space components of the kaon vector
field \cite{Fuchs98}. However, the calculations of Ref. \cite{David99},
where the spatial components have also been taken into account, have produced 
a larger negative kaon flow, which disagrees with Ref. \cite{Fuchs98}.
The latest calculations by the T\"ubingen QMD group \cite{Zheng04} have
corroborated their earlier analysis \cite{Fuchs98} and demonstrated
that the new FOPI data on the kaon in-plane flow \cite{Herr99} are
best described by using the kaon potential given by the Brown-Rho (BR)
parametrization \cite{BR96} ($U_K(\rho_0) \simeq 30$ MeV).
At $\rho \leq \rho_0$ the kaon potential in the BR parametrization is close 
to the one in the impulse approximation(c.f. Refs. \cite{Li951,SMB97}).
Recent self-consistent calculations of Refs. \cite{Korpa04,Tolos05}
show even stronger repulsion for kaons 
($U_K(\rho_0) = 36$ MeV in Ref. \cite{Korpa04} and 
$U_K(\rho_0) = 39$ MeV in Ref. \cite{Tolos05}).

The $K^\pm$ azimuthal distributions at midrapidity (squeeze-out) have also 
been studied within an earlier version of the T\"ubingen QMD model in Ref. 
\cite{Wang99}. It has been concluded in \cite{Wang99}, that the $K^+$ 
squeeze-out is caused mainly by the repulsive $K^+$ potential. However, the 
quantitative agreement with experimental data for Au+Au at 1 A GeV 
\cite{Shin98} was achieved only with the static kaon potential in 
\cite{Wang99}. The Lorentz force, as has been concluded in \cite{Wang99}, 
destroys the agreement with the data by reducing the $K^+$ squeeze-out signal 
strongly (we will come back to this point later on).

The present work is an attempt to describe the $K^+$ data on multiplicities 
and phase space distributions at the beam energies of 1-2 A GeV 
\cite{Sturm01,Shin98,Rit95,Herr99,Menzel00,Crochet00,Forst03,Uhlig04}
on the basis of a BUU model \cite{EBM99,MEPhD,LM03}. 
The BUU model \cite{EBM99,MEPhD,LM03} includes a large set of the baryonic
resonances (see the next Section). Due to the channels
$N R \to K Y N$ and $N R \to K \bar{K} N N$, where $R$ stands for a
nonstrange baryon resonance, the resonances enhance strangeness production. 
Thus, despite of many previous transport theoretical
studies on the strangeness production at SIS energies 
(c.f. Refs. \cite{AK85,Fuchs01,Li95,Li951,Li96,Li961,Brat97,David99,
Hart03,Hart031,Mish04,Zheng04}) it would be interesting 
to confront also our calculations with experimental data, since, e.g., 
the QMD models 
\cite{Aich91,David99,Fuchs01,Hart03,Hart031,Zheng04,QMD} propagate only 
$\Delta(1232)$ and $N^*(1440)$ resonances and the HSD model 
\cite{Brat97,CB99,Mish04} propagates only $\Delta(1232)$, $N^*(1440)$ and 
$N^*(1535)$ resonances (see also Ref. \cite{THW04} for the comparison
of the different transport calculations).

The main purpose of our study is to clarify whether a kaon potential 
is actually needed to describe the data and, if so, how strong it must be. 
In particular, we will analyse the recent KaoS data \cite{Uhlig04} 
on the kaon azimuthal distributions. 

In Sect. II a brief description of the BUU model is given.
Sect. III contains the results of numerical calculations. In Sect. IV we 
summarize our results and draw some conclusions.

\section{ The BUU model }

Our calculations are based on the BUU model in the version described in
Refs. \cite{EBM99,MEPhD,LM03}. The model explicitly propagates all 
$N^*$ and $\Delta$ resonances that are rated with at least 2 stars 
in the analysis of Ref. \cite{Manley92}, which includes the $N^*$ states 
$P_{11}(1440)$, $D_{13}(1520)$, $S_{11}(1535)$, $S_{11}(1650)$, 
$D_{15}(1675)$, $F_{15}(1680)$, $P_{13}(1879)$, $F_{17}(1990)$,
$G_{17}(2190)$ and the $\Delta$ states $P_{33}(1232)$, $P_{33}(1600)$, 
$S_{31}(1620)$, $S_{31}(1900)$, $F_{35}(1905)$, $P_{31}(1910)$, 
$D_{35}(1930)$, $F_{37}(1950)$, $D_{35}(2350)$.
Also the $S=-1$ baryons $Y=\Lambda(1116)$, $\Sigma(1189)$ and 
$Y^*=S_{01}(1405)$,
$D_{03}(1520)$, $P_{01}(1600)$, $S_{01}(1670)$, $D_{03}(1690)$,
$S_{01}(1800)$, $P_{01}(1810)$, $F_{05}(1820)$, $D_{05}(1830)$,
$P_{03}(1890)$, $G_{07}(2100)$, $F_{05}(2110)$,
$P_{13}(1385)$,
$P_{11}(1660)$, $D_{13}(1670)$, $S_{11}(1750)$, $D_{15}(1775)$,
$F_{15}(1915)$, $F_{17}(2030)$ are propagated explicitly in the model.
The  $Y^*$-resonances are the intermediate states of the strangeness
exchange reactions $\pi Y \leftrightarrow Y^* \leftrightarrow \bar K N$.

In the meson sector, the following particles are propagated:
$\pi$, $\eta$, $\rho$, $\sigma$, $\omega$, $\eta^\prime$,
$\phi$, $\eta_c$, $J/\psi$, $K$, $\bar K$, $K^*$, $\bar K^*$.
Also the cascades, charmed baryons and mesons are propagated,
which are, however, irrelevant degrees of freedom at SIS energies.

The model has been successfully applied to $\gamma$ and $e^-$ induced
reactions on nuclei \cite{EBM99,Lehr00} and to the description of the
collective nucleon flows in heavy-ion collisions at 0.1-2 A GeV
\cite{LCGM00}. The reproduction of the pion abundancies in heavy-ion 
collisions at SIS energies requires, however, to apply the in-medium 
reduced resonance production/absorption cross sections 
$N N \leftrightarrow N R$ \cite{LM03,LCLM01}. We will drop the detailed
description of the model which can be found in Refs. \cite{EBM99,MEPhD,LM03}
concentrating here only on some novel features related to strangeness
production.

For kaon production in baryon-baryon collisions we include the
following channels: $N N \to N Y K$, $N N \to \Delta Y K$, 
$\Delta N \to N Y K$, $\Delta N \to \Delta Y K$ and 
$\Delta \Delta \to \Delta Y K$, where $\Delta \equiv P_{33}(1232)$
and $Y = \Lambda(1115)$ or $\Sigma(1189)$, with all possible isospin
combinations of the incoming and the outgoing particles. The channel
$\Delta \Delta \to N Y K$ is not included, since it is possible only
through an exchange by an on-shell pion \cite{Tsush99}, which is
already included in BUU via the consecutive $\Delta \leftrightarrow N \pi$ 
and $\pi B \to K Y$ processes, where $B \equiv N$ or $R$. The isospin-dependent
parametrizations for the cross sections of kaon production 
in the $NN, N\Delta$ and $\Delta\Delta$ collisions are taken from 
Ref. \cite{Tsush99}. In addition, we allow to produce a kaon-hyperon pair
in a collision between any two nonstrange baryons: cross sections of
these processes are obtained from the cross sections of Ref. \cite{Tsush99}
by replacing an incoming $N^*$ resonance by the nucleon and an incoming
higher $\Delta$ resonance by the $\Delta(1232)$ for the same 
$\sqrt{s}$. The $K \bar K$ pair production in a baryon-baryon collision
$B B \to N N K \bar K$ is also included via cross sections parametrized
in Ref. \cite{Cas97}. 

The pion-baryon collisions $\pi B \to Y K$  also contribute quite 
significantly to the strangeness production in heavy-ion collisions.
If the incoming baryon $B$ is a nucleon or a 
$\Delta(1232)$-resonance, the cross sections from Ref. \cite{Tsush97}
are applied. 
If $B=N^*$ or higher $\Delta$, it is substituted by the nucleon
or $\Delta(1232)$-resonance, depending on the isospin, and then the 
corresponding cross sections from Ref. \cite{Tsush97} are used. 
The $\pi B \to N K \bar K$ process is taken into account with a cross section 
parametrized in Ref. \cite{Sib97}.
The kaon elastic, including charge exchange, and inelastic scattering 
processes $K N \to K N$ and $K N \to K N \pi$ 
are also taken into account: their cross sections are fitted to the data 
\cite{Bald87} (see Ref. \cite{MEPhD}).

Since in the present work we concentrate on the kaon production at 
energies below 2 A GeV, the kaon production channels $B B \to N Y K$ and 
$\pi B \to Y K$ are those relevant for our study. 
The reactions including antikaons play, practically, no role
for kaon production (see Table~\ref{tab:channels} below). 
The kaon production is treated perturbatively and the FRITIOF mechanism is
switched off in this work, since the kaon multiplicity per nucleus-nucleus 
collision is still quite small at the considered beam energies 
(see Fig.~\ref{fig:rateint} below).

In our calculations the nucleons are propagated in a Skyrme-like mean
field including the momentum-dependent part (c.f. Ref. \cite{LCGM00}).
Most of the calculations are done with a soft momentum-dependent
mean field (SM, $K=220$ MeV). When, for comparison, also a hard 
momentum-dependent mean field (HM, $K=380$ MeV) is used, this
will be mentioned in the text below. It is assumed that all the
nonstrange baryonic resonances experience the same mean field as nucleons.
For the hyperons, according to the fraction of the nonstrange quarks,
we apply the nucleon mean field multiplied by 2/3 \cite{Fuchs01,Fang94,CB99}.

The $K^\pm$ single-particle energies are expressed as 
\begin{equation}
   \omega_K^\pm({\bf k}) = \pm V^0 + \sqrt{{\bf k}^{*2} + m_K^{*2}}~,
                                                  \label{omegavsk}
\end{equation}
${\bf k}^* = {\bf k} \mp {\bf V}$ is the kaon kinetic momentum,
$V^\mu = (V^0,{\bf V})$ is the kaon vector field and $m_K^*$ is the kaon
effective (Dirac) mass. According to Refs. \cite{BR96,Zheng04}, the kaon 
effective mass and the kaon vector field are given by the following 
expressions~:
\begin{eqnarray}
  m_K^* & = & \sqrt{ m_K^2 - {\Sigma_{KN} \over f_\pi^2}\rho_s + V^2 }~,
                                                  \label{mkst} \\
  V^\mu & = & {3 \over 8 f_\pi^{*2}} j^\mu~,         \label{Vmu}
\end{eqnarray}
where $m_K=0.496$ GeV is the bare kaon mass, $\rho_s$ and $j^\mu$ are
the baryon scalar density and the four-current, respectively.
The parameters which appear in Eqs. (\ref{mkst}),(\ref{Vmu}) are 
$\Sigma_{KN}=0.450$ GeV --- the kaon-nucleon sigma term, 
$f_\pi = 0.093$ GeV --- the vacuum pion decay constant and 
$f_\pi^* = \sqrt{0.6} f_\pi$ --- the in-medium pion decay constant at
normal nuclear matter density $\rho_0$ \cite{BR96}.  

Using the in-medium pion decay constant rather than the vacuum one 
in Eq. (\ref{Vmu}) leads to the desirable consequence \cite{BR96}, that
the resulting kaon vector potential (\ref{Vmu}) is just 1/3 of the nucleon 
vector potential given by the relativistic mean field model. In the
scalar term $\propto \rho_s$ (Eq. (\ref{mkst})) the vacuum pion decay
constant is used, since for kaons the higher order (range) term 
in the chiral expansion cancels an effect of the in-medium pion
decay constant in this case \cite{BR96}. The cancellation, however, does
not take place for the antikaons \cite{BR96}. Thus, Eq.(\ref{mkst})
can be considered for the antikaons as a phenomenological parametrization,
which, nevertheless, leads to reasonable values of the $\bar K$ potential
(see below).

The kaon (antikaon) potential $U_{K(\bar K)}$ is defined as
\begin{equation}
   U_{K(\bar K)}({\bf k}) 
 = \omega_K^\pm({\bf k}) -  \sqrt{{\bf k}^2 + m_K^2}~. \label{UK}
\end{equation}
Following Ref. \cite{Zheng04}, we will denote the kaon (antikaon) potential 
with parameters determined above as the BR potential which 
will be used as the default potential in our calculations. 
However, as in \cite{Zheng04}, we will perform for a comparison also
some calculations applying the kaon (antikaon) potential with 
$\Sigma_{KN}=0.350$ GeV and with the free pion decay constant $f_\pi$ instead 
of $f_\pi^*$ in the vector potential (\ref{Vmu}) which we call a Ko-Li (KL) 
potential \cite{Li951}.   

\begin{figure}

\includegraphics{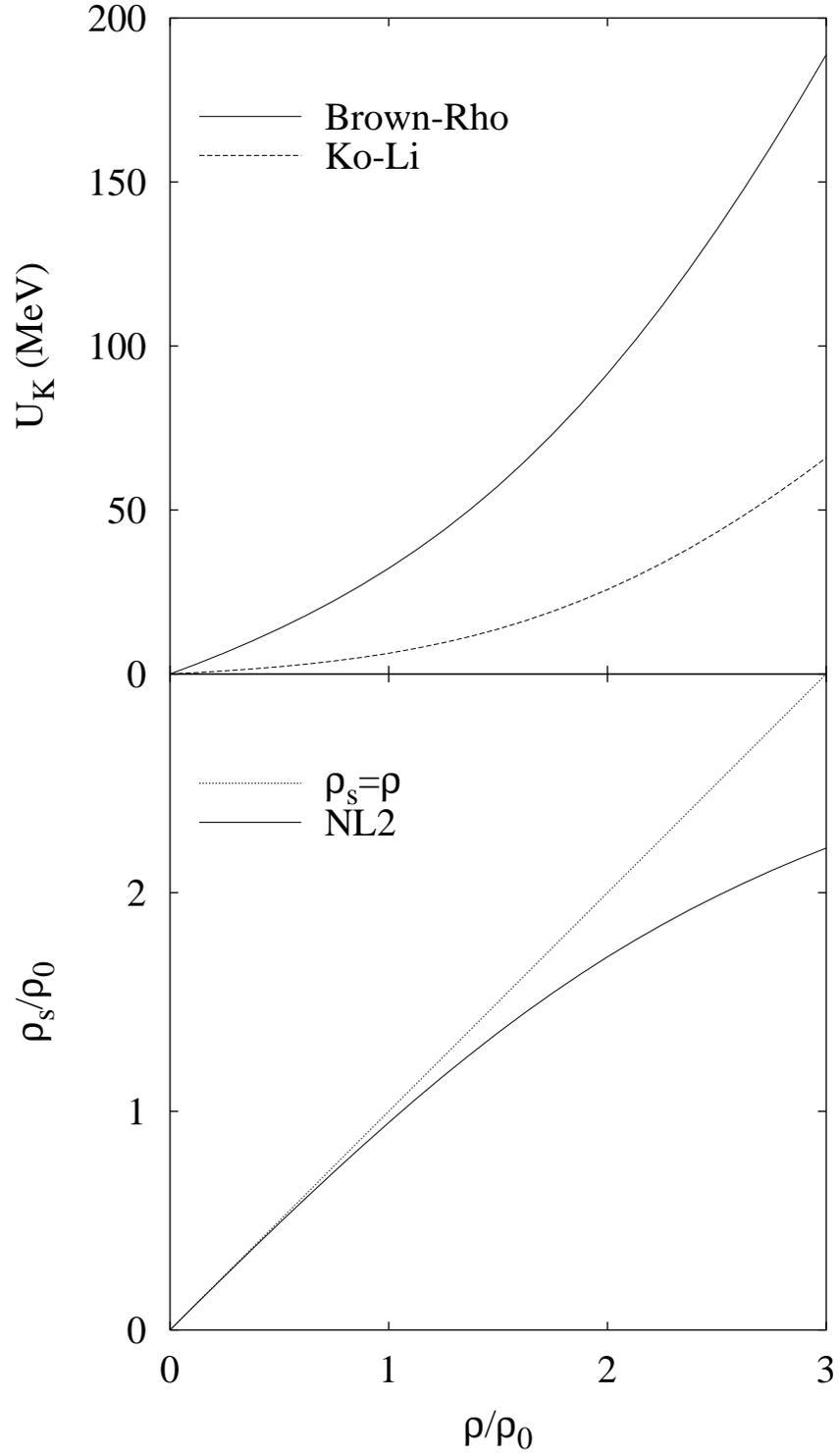}

\caption{\label{fig:ukaon} Upper panel: the kaon potential 
at zero momentum as a function of the nuclear matter density.
Lower panel: the scalar density vs the nuclear matter density.}

\end{figure}

Both choices of the kaon potential, BR and KL, are shown for the static case 
(${\bf k}=0$) in the upper panel of Fig.~\ref{fig:ukaon} as functions of the 
baryon density. The BR potential is strongly repulsive ($U_K(\rho_0)=32$ MeV),
while the KL potential is much weaker ($U_K(\rho_0)=6$ MeV). 
We have to point out here, that our
potentials are slightly more repulsive at higher density than the
corresponding potentials from Ref. \cite{Zheng04} due to different
versions of the relativistic mean field model used to evaluate the
scalar density. For an orientation, we show in the lower
panel of Fig.~\ref{fig:ukaon} the scalar density as a function of
the baryon density which is given by the NL2 model \cite{Lee86} 
applied in our calculations. In the case of antikaons, the BR (KL)
parametrization produces  $U_{\bar K}(\rho_0)=-144$ MeV (-100 MeV).

In the BUU implementation we solve the Hamiltonian equations of motion
for the kaon (antikaon) test particles, where the Hamilton function is the 
single-particle energy (\ref{omegavsk}) which now implicitly depends
also on space and time via the vector field $V^\mu({\bf r},t)$ 
and the effective mass $m^*_K({\bf r},t)$~:
\begin{eqnarray}
    \dot{\bf r} & = &  {\partial \omega_K^\pm({\bf k},{\bf r},t) 
                         \over \partial {\bf k}}
                  = {{\bf k}^* \over E^*}~,           \label{Hameq1} \\
    \dot{\bf k} & = & -{\partial \omega_K^\pm({\bf k},{\bf r},t)
                         \over \partial {\bf r}}
= -{m_K^* \over E^*}{\partial m_K^* \over \partial {\bf r}}
   \mp {\partial V_0 \over \partial {\bf r}}
      \pm {k_\alpha^* \over E^*} {\partial V_\alpha \over \partial {\bf r}}~,
                                                       \label{Hameq2}
\end{eqnarray} 
where $E^* = \sqrt{{\bf k}^{*2} + m_K^{*2}}$.
These equations of motions are completely equivalent to those 
in the covariant form derived in Refs. \cite{Fuchs98,Zheng04}.
The last term in the r.h.s. of Eq.(\ref{Hameq2}) is a velocity-dependent
(Lorentz-like) force caused by the spatial components ${\bf V}$ of the
kaon vector field.

Potentials shift the particle production thresholds in nuclear medium.
We take this effect into account, following Ref. \cite{Fuchs01}, by
replacing the bare center-of-mass (c.m.) energy $\sqrt{s_{free}}$ in the
argument of the cross sections 
$\sigma_{BB \to {\rm strangeness}}(\sqrt{s_{free}})$ and 
$\sigma_{\pi B \to {\rm strangeness}}(\sqrt{s_{free}})$ by a corrected
quantity. E.g., for the process $N N \to N Y K$ we replace $\sqrt{s_{free}}$
by $\sqrt{s} - \tilde m_N - \tilde m_Y - \tilde m_K + m_N + m_Y + m_K$,
where $\sqrt{s}$ is the total in-medium c.m. energy of colliding particles
including their mean field potentials, 
$\tilde m_X \equiv \varepsilon_X({\bf p}_X^{c.m.}=0)$ ($X=N,~Y,~K$) are
the in-medium masses of the particles defined as their energies at rest
in the c.m. frame of colliding particles \cite{com1}. 

In the simulation of a three-body phase space for the outgoing particles
$B Y K$ ($B=N$ or $\Delta(1232)$ here) we use the kaon momentum distribution
in the $B Y K$ c.m. frame proposed in Refs. \cite{LLB97,Zheng04}~:
\begin{equation}
   dW_K \simeq \left({p \over p_{\rm max}}\right)^3
               \left(1-{p \over p_{\rm max}}\right)^2~,    \label{dWK}
\end{equation}
where $p_{\rm max} = [(s - (\tilde m_B + \tilde m_Y)^2 + \tilde m_K^2)^2
/4s - \tilde m_K^2]^{1/2}$ is the maximal kaon momentum in the $B Y K$ c.m. 
frame. The angular distribution of the produced kaon is chosen in the
empirical form \cite{Zheng04,Sturm_thesis}~:
$d\sigma/d\cos\Theta_{c.m.} \propto (1+a\cos^2\Theta_{c.m.})$ 
with $a=1.2$ \cite{Sturm_thesis}. These two modifications lead to a softer 
kaon $p_{\rm lab}$ spectrum with respect to the simulation using an ideal
Dalitz 3-body decay \cite{Zheng04}.

As it was demonstrated in Ref. \cite{LM03}, taking into account the Dirac
masses of the baryons reduces the cross sections $NN \leftrightarrow NR$
in nuclear matter strongly, which brings our BUU calculations in a better
agreement with the experimental data on pion multiplicities. In the
calculations of the present work, if opposite is not stated explicitly, 
we use the in-medium cross sections $NN \leftrightarrow NR$ and 
$NN \leftrightarrow NN\pi$ \cite{LM03} computed with the Dirac masses from 
the NL2 model \cite{Lee86}. The exchange pion collectivity effect and vertex 
corrections in the $NN \leftrightarrow N\Delta(1232)$ cross sections included
in the calculations of Ref. \cite{LM03} are neglected in the present work
for simplicity. For brevity, we will call below a calculation employing
the SM nucleon mean field, the BR kaon potential and the in-medium 
$NN \leftrightarrow NR$ cross sections the standard one.

\section{ Numerical results }

\subsection{ Kaon production channels }

\begin{figure}

\includegraphics{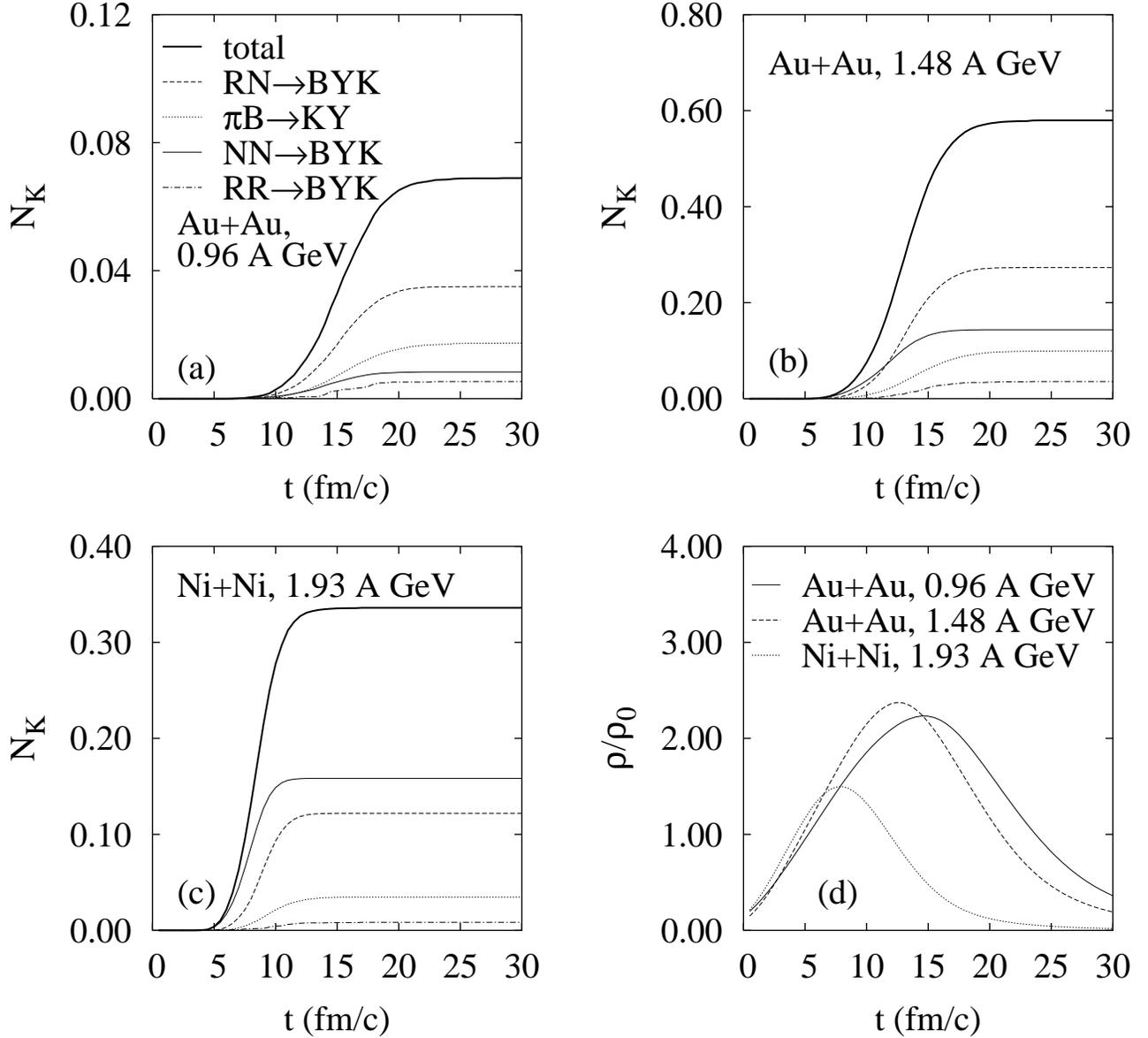}

\caption{\label{fig:rateint} Number of kaons ($K^+$'s and $K^0$'s) 
as a function of time for central ($b=0$ fm) collisions
Au+Au at 0.96 A GeV (a), Au+Au at 1.48 A GeV (b) and Ni+Ni at 1.93 A GeV (c) 
for the different production channels: 
$R N \to B Y K$ --- dashed lines, $\pi B \to K Y$ --- dotted lines,
$N N \to B Y K$ --- thin solid lines and $R R \to B Y K$ --- dash-dotted
lines. Total kaon number is represented by
thick solid lines in the panels a,b,c. Panel d shows the time evolution of a 
central baryon density for the central collisions Au+Au at 0.96 A GeV --- 
solid line, Au+Au at 1.48 A GeV --- dashed line and Ni+Ni at 1.93 
A GeV --- dotted line.}

\end{figure}

Before starting the comparison with experimental data, we have looked at the 
time evolution of the kaon number produced by the different channels, 
which is displayed in Fig.~\ref{fig:rateint}a,b,c for the central
collisions Au+Au at 0.96 A GeV, Au+Au at 1.48 A GeV and Ni+Ni at 1.93 A GeV,
respectively. Contributions of the channels to the total kaon number are
summarized in Table~\ref{tab:channels}. In Fig.~\ref{fig:rateint} and
in Table ~\ref{tab:channels} ``$B$'' in the initial state denotes 
$N$ or $R$, while ``$B$'' in the final state denotes $N$ or $\Delta(1232)$.
The main kaon production channels are $R N \to B Y K$, $N N \to B Y K$
and $\pi B \to K Y$. Other channels contribute all together on the level 
of $\sim 10$\%.

At the lowest energy of 0.96 A GeV the direct channel $N N \to N Y K$ 
is deeply subthreshold and, thus, its contribution to the total kaon
number is small. With increasing beam energy the contribution of the 
$N N \to N Y K$ channel grows quickly, so that at 1.93 A GeV this channel
becomes already the main one.  The $R N \to B Y K$ channel dominates
at 0.96 and 1.48 A GeV. It is also the second important at 1.93 A GeV.
 
The $\pi B \to K Y$ channel is the second important at 0.96 A GeV,
but its contribution decreases with increasing energy.
This is quite natural, since a channel caused by secondary particles
is energetically favourable at subthreshold energies.
Another reason is that the pions are produced mainly in resonance decays
$R \to N \pi$. The time scale of a nucleus-nucleus collision gets
shorter with increasing beam energy. Therefore, more and more resonances
will decay at the final low baryon density stage, when the probability
of the pion-baryon collisions is small.

By comparing the time evolution of the kaon production 
(Fig.~\ref{fig:rateint}a,b,c) and of the central baryon density
(Fig.~\ref{fig:rateint}d) we see that the largest rate of the kaon
production takes place at about the time of the maximum compression.
Thus, the kaon production delivers a signal from the highest compression
stage of a heavy-ion collision without much distortion from the later
stage, since kaons are not absorbed in the nuclear medium.
This important kaon property has given, in particular, an opportunity
to determine the nuclear EOS from the kaon yields \cite{AK85,Fuchs01}.
 
\subsection{ Kaon multiplicities and spectra }

\begin{figure}

\includegraphics{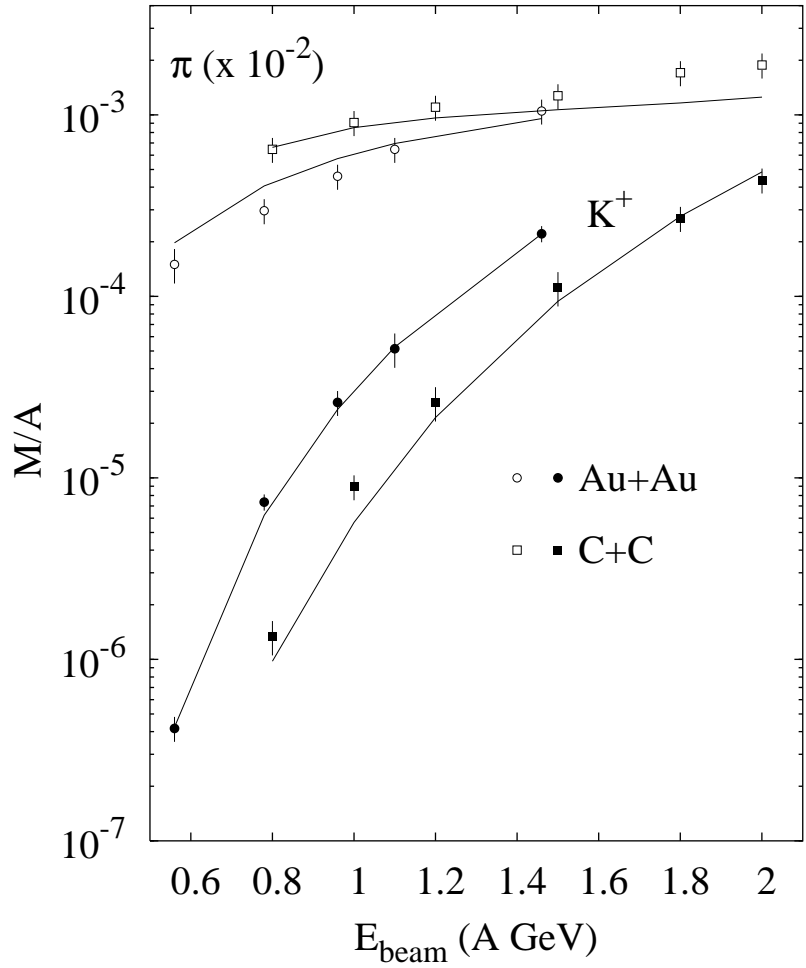}

\caption{\label{fig:auaucc_mult} Pion (upper two lines) and kaon
(lower two lines) multiplicities per projectile nucleon in the C+C
(highest and lowest lines) and in the Au+Au (two lines in the middle)
systems vs the projectile energy. Pion multiplicity includes all pions 
and is scaled by a factor of $10^{-2}$. Data are from Ref. 
\cite{Sturm01}.}

\end{figure}

First, we check whether our model is able to reproduce the pion and kaon
total multiplicities. Fig.~\ref{fig:auaucc_mult} shows the inclusive
multiplicities of $\pi$'s ($\pi = \pi^- + \pi^0 + \pi^+$) and K$^+$'s 
per projectile nucleon for the systems Au+Au and C+C as functions of
the beam energy in comparison to the data from Ref.~\cite{Sturm01}.
In the calculations the particle multiplicities were impact parameter
weighted in the region $b < 14$ fm for the Au+Au collisions and
in the region $b < 5$ fm for the C+C collisions which corresponds to
the geometrical cross sections. The pion multiplicity for the C+C
system is well reproduced, except for the points at 1.8 and 2.0 A GeV,
where we underpredict the data due to neglecting the string (FRITIOF) 
mechanism of particle production. In the case of Au+Au collisions
our calculation produces too many pions at lower beam energies
in spite of using the in-medium reduced $NN \leftrightarrow NR$
cross sections. Applying faster dropping Dirac masses of the baryons
with nuclear density, e.g. given by the NL1 model \cite{Lee86}, as well
as taking into account the pion collectivity and vertex corrections
in the $NN \leftrightarrow N\Delta(1232)$ cross sections \cite{LM03} would 
lead to a better description of the pion multiplicity in the Au+Au case,
however, at the cost of too low pion multiplicity for C+C collisions. 
The K$^+$ multiplicities are rather well reproduced for both systems 
Au+Au and C+C in the full SIS energy range.

Fig.~\ref{fig:d2sig_au096au} shows the laboratory momentum kaon spectra
for the Au+Au collisions at 0.96 A GeV in comparison to the data 
\cite{Sturm01}. The calculation without kaon potential 
(Fig.~\ref{fig:d2sig_au096au}a, dashed lines with open up triangles) 
overestimates the kaon production strongly. The calculations, which include 
the kaon potential (Fig.~\ref{fig:d2sig_au096au}a, BR --- solid lines with open
circles, KL --- dotted lines with open squares) lead to a reduced lower 
momentum part of the spectra in better agreement with the data. 
The high momentum tail of the K$^+$ spectrum still remains too high in 
the calculations with the kaon potential.

In Fig.~\ref{fig:d2sig_au096au}b we compare our standard calculation with
the calculation employing the vacuum $NN \leftrightarrow NR$ cross sections.
We see that dropping the in-medium corrections to the $NN \leftrightarrow NR$
cross sections results in about 50\% larger kaon production cross section.

In order to see the origin of the produced kaons, we performed in 
Fig.~\ref{fig:d2sig_au096au}c a channel decomposition of the laboratory
momentum spectrum \cite{com4}. The largest contribution is provided by the 
pion-baryon collisions ($\pi B$). The nucleon-$\Delta(1232)$ and 
nucleon-higher resonance channels have also big contributions, comparable to 
the $\pi B$ channel. This explains also a sensitivity of the kaon production 
cross section to the choice of the $NN \leftrightarrow NR$ cross sections 
(see Fig.~\ref{fig:d2sig_au096au}b). The nucleon-nucleon collisions contribute
only about 10\% to the total kaon production cross section, since, 
at $E_{lab}=0.96$ A GeV, they are mostly below the kaon production threshold.

In Fig.~\ref{fig:d2sig_au096au}d we present a calculation with the HM nucleon
mean field, which produces less K$^+$'s (dashed line with open diamonds) 
than the standard calculation employing the SM mean field (solid line with
open circles). The pion off-shellness effect \cite{LM02} (dotted line with 
open squares) does not increase the K$^+$ multiplicity strong enough to bring 
the calculation with the HM mean field to agreement with the data.

\begin{figure}

\includegraphics{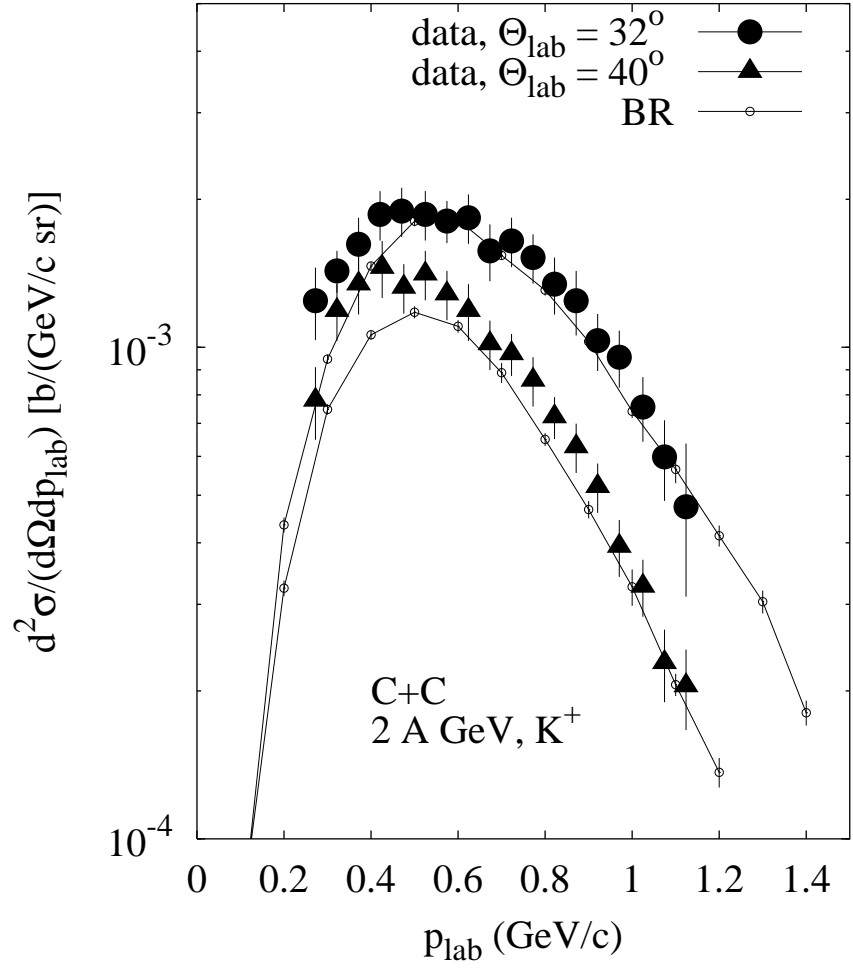}

\caption{\label{fig:d2sig_c200c} Inclusive K$^+$ production cross section 
as a function of the laboratory momentum for C+C at 2 A GeV at 
$\Theta_{\rm lab} = 32^o$ and $40^o$. Data are taken from Ref. 
\cite{Sturm_thesis}.}

\end{figure}

Fig.~\ref{fig:d2sig_c200c} shows the laboratory momentum kaon spectra for
the system C+C at 2 A GeV in comparison with the data from Ref. 
\cite{Sturm_thesis}. It is interesting that for this light system we reproduce
rather well the slopes of the experimental spectra. A similar result was
obtained earlier in Ref. \cite{Zheng04} in a calculation with an anisotropic 
angular distribution of the produced kaon in a NN collision. 

\begin{figure}

\includegraphics{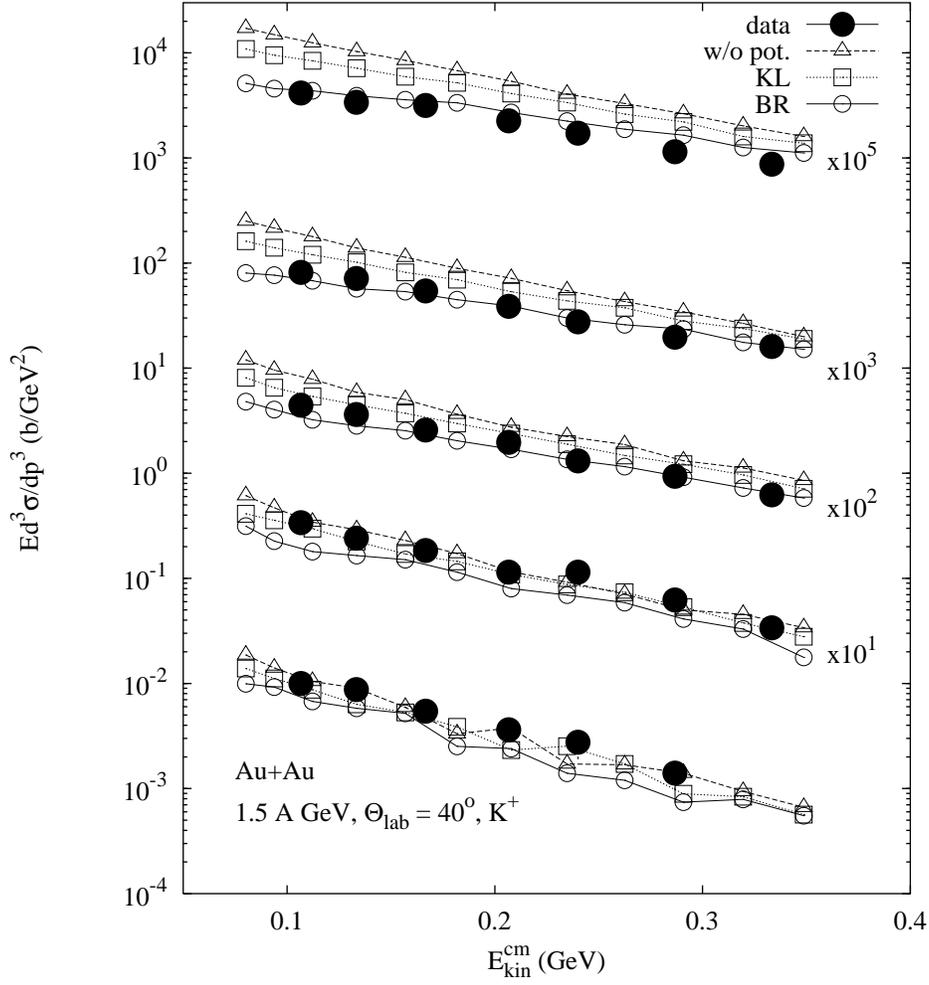}

\caption{\label{fig:d3sig_au150au_kp} Differential K$^+$ production
cross section at $\Theta_{\rm lab} = 40^o$ from Au+Au collisions 
at 1.5 A GeV for different centrality bins as a function of the kinetic 
energy in the c.m. system. Data points are taken from Ref. \cite{Forst03}
and correspond to the following centrality bins (from top to bottom
with decreasing centrality): 5\%, 15\%, 15\%, 25\% and 40\% of the
reaction cross section. Corresponding calculations are performed within
the following impact parameter regions (from the upper to lower lines):
$b\leq3$ fm, $b=4-6$ fm, $b=7-8$ fm, $b=9-10$ fm and $b=11-14$ fm.
Spectra are scaled by the factors $10^5, 10^3, 10^2, 10^1, 10^0$ from top
to bottom.}

\end{figure}

Fig.~\ref{fig:d3sig_au150au_kp} shows the c.m. kinetic energy spectra
of $K^+$'s from Au+Au collisions at 1.5 A GeV for various event centrality
classes. For the most central collisions, the calculation without potential 
(upper dashed line with open triangles) clearly overestimates kaon yield and 
also has a too steep slope. Including kaon potentials (BR --- solid lines
with open circles, KL --- dotted lines with open squares) reduces the kaon 
yield and reduces the steepness of the slope inproving, thus, an agreement 
with the data for central events. With decreasing centrality the slopes 
of the spectra calculated with and without potentials get similar. 
This is expected since in peripheral events the compression
is less and, therefore, any influence of the kaon potential is reduced.
For the most central events we observe that the calculation employing the
BR potential describes the data best, while for the peripheral 
events the calculation without potential is in best agreement with the 
data.  

The qualitative trends presented in Fig.~\ref{fig:d3sig_au150au_kp} can 
be better visible if one fits the spectra as follows \cite{Forst03}:
$E d^3\sigma/ dp^3 = C \cdot E \cdot \exp(-E/T)$, where $E=E_{kin}^{cm}+m_K$
is the total energy of a kaon in the c.m. system.
In the upper panel of Fig.~\ref{fig:slope_au150au_kp} we show the inverse 
slope parameter $T$ of the $K^+$ c.m. kinetic energy spectra as a function 
of a participant number $A_{part}$ for the Au+Au system at 1.5 A GeV.
The participant number was determined for each impact parameter from the 
geometrical overlap of colliding nuclei assuming sharp nuclear surfaces.
As we already saw in Fig.~\ref{fig:d3sig_au150au_kp}, the difference
between the inverse slope parameters from the calculations with and without
kaon potentals is small at the peripheral collisions and increases with
the collision centrality. The calculation with the KL potential provides
the best description of the inverse slope parameter for all $A_{part}$. 

The lower panel of Fig.~\ref{fig:slope_au150au_kp} shows the
$K^+$ multiplicity per participating nucleon as a function of $A_{part}$.
The calculation without potential is above the data by a factor of two. 
The KL potential is not strong enough to get the correct multiplicities 
\cite{com2}. Only using the BR potential reduces the kaon multiplicity 
to a good agreement with the data.

\begin{figure}

\includegraphics{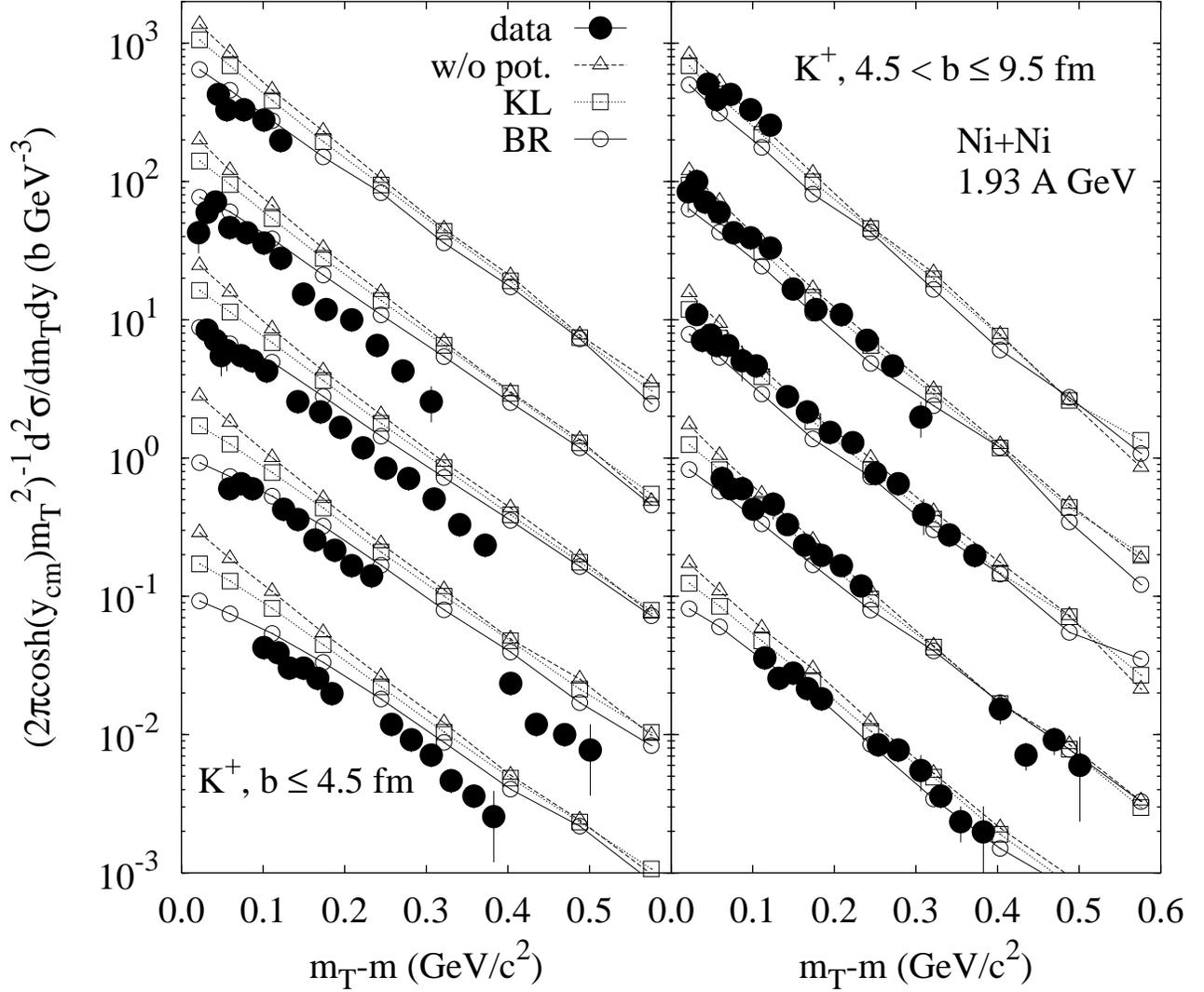}

\caption{\label{fig:dNdmt_ni193ni_kp} Transverse mass spectra of $K^+$'s
from the Ni+Ni collisions at 1.93 A GeV for central events (left panel)
and peripheral events (right panel) in comparison to the data
from Ref. \cite{Menzel00}.  The spectra are extracted in the following
c.m. rapidity intervals (from top to bottom): $-0.69 < y_{cm} < -0.54$,
$-0.54 < y_{cm} < -0.39$, $-0.39 < y_{cm} < -0.24$, $-0.24 < y_{cm} < -0.09$,
$-0.09 < y_{cm} < 0.06$. The scaling factors $10^4, 10^3, 10^2, 10^1$ and 
$10^0$ are applied to the spectra from top to bottom.}

\end{figure}

\begin{figure}

\includegraphics{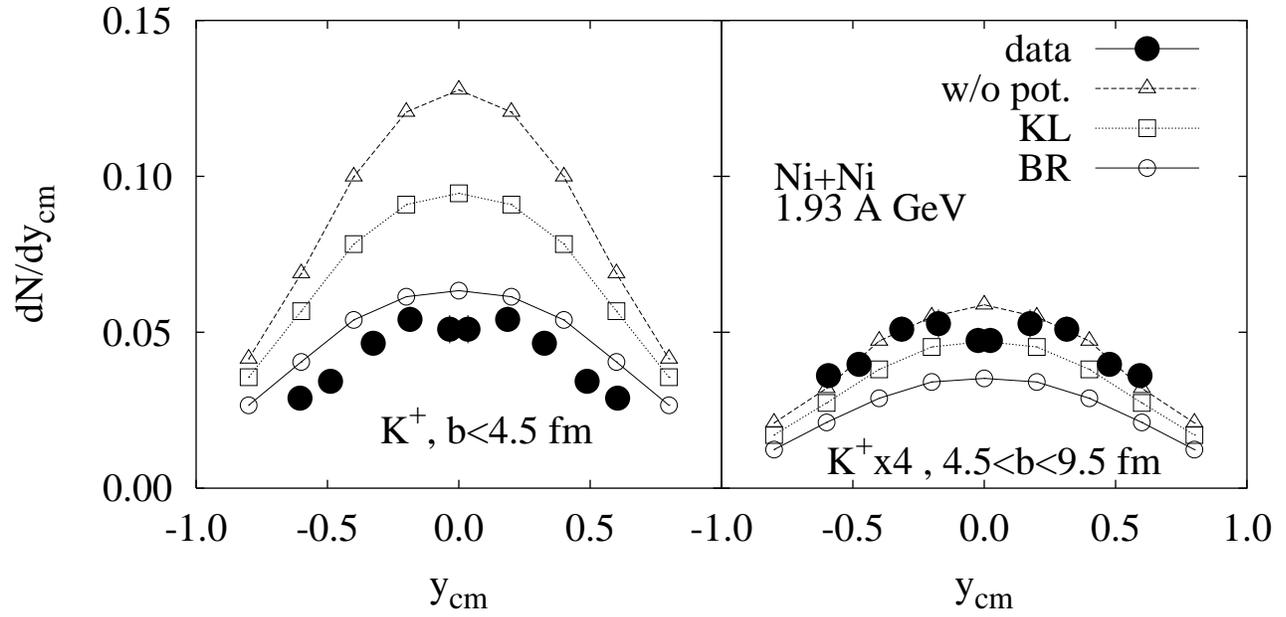}

\caption{\label{fig:dNdy_ni193ni_kp} $K^+$ c.m. rapidity distributions
from the central (left panel) and peripheral (right panel) collisions
of Ni+Ni at 1.93 A GeV in comparison to the the data from Ref. \cite{Menzel00}.
The distributions for the peripheral collisions are multiplied by a factor 
of 4.}

\end{figure}

Figs.~\ref{fig:dNdmt_ni193ni_kp} and \ref{fig:dNdy_ni193ni_kp} show,
respectively, the $K^+$ transverse mass ($m_T \equiv \sqrt{p_T^2+m^2}$) and 
c.m. rapidity spectra for Ni+Ni collisions at 1.93 A GeV in comparison to 
the data \cite{Menzel00}.  
As in the case of Au+Au collisions at 1.5 A GeV we observe that the 
calculation without kaon potential overpredicts the kaon yields
and produces too steep slopes at low $m_T$'s for the central collisions
(left panels of Figs.~\ref{fig:dNdmt_ni193ni_kp} and 
\ref{fig:dNdy_ni193ni_kp}).
The BR potential provides the most reasonable 
description of the data for the central collisions. 
However, the steepness of the slopes is somewhat
underestimated by the calculation with the BR potential.
The slopes are, again, best described by the calculation with the KL 
potential. In the case of peripheral collisions (right panels of 
Figs.~\ref{fig:dNdmt_ni193ni_kp} and \ref{fig:dNdy_ni193ni_kp})
we observe that the calculations without potential and with the KL
potential produce an equally good description of the data, while
the BR potential leads to a slight underestimation of the kaon yield.

\subsection{ Collective flows }

Extraction of collective flows requires a knowledge of a reaction plane,
i.e. the plane which is parallel to the impact parameter and to the
beam momentum \cite{DO85}. In theoretical calculations the reaction plane is
obviously known. Experimentally, however, the reaction plane
is usually taken parallel to the difference of transverse momenta of charged 
particles --- which are mostly protons at SIS energies --- in the forward and 
backward hemispheres in the c.m. frame of colliding nuclei.
Thus, a collective flow carries an information about the correlations
between the particle under study and the remaining protons. This makes the 
collective flow observables extremely useful to constrain the mean field 
potentials of the particles (c.f. Refs. \cite{Li95,Li951,Li96,Li961,Brat97} 
for $K$ and $\bar K$ flows).

\begin{figure}

\includegraphics{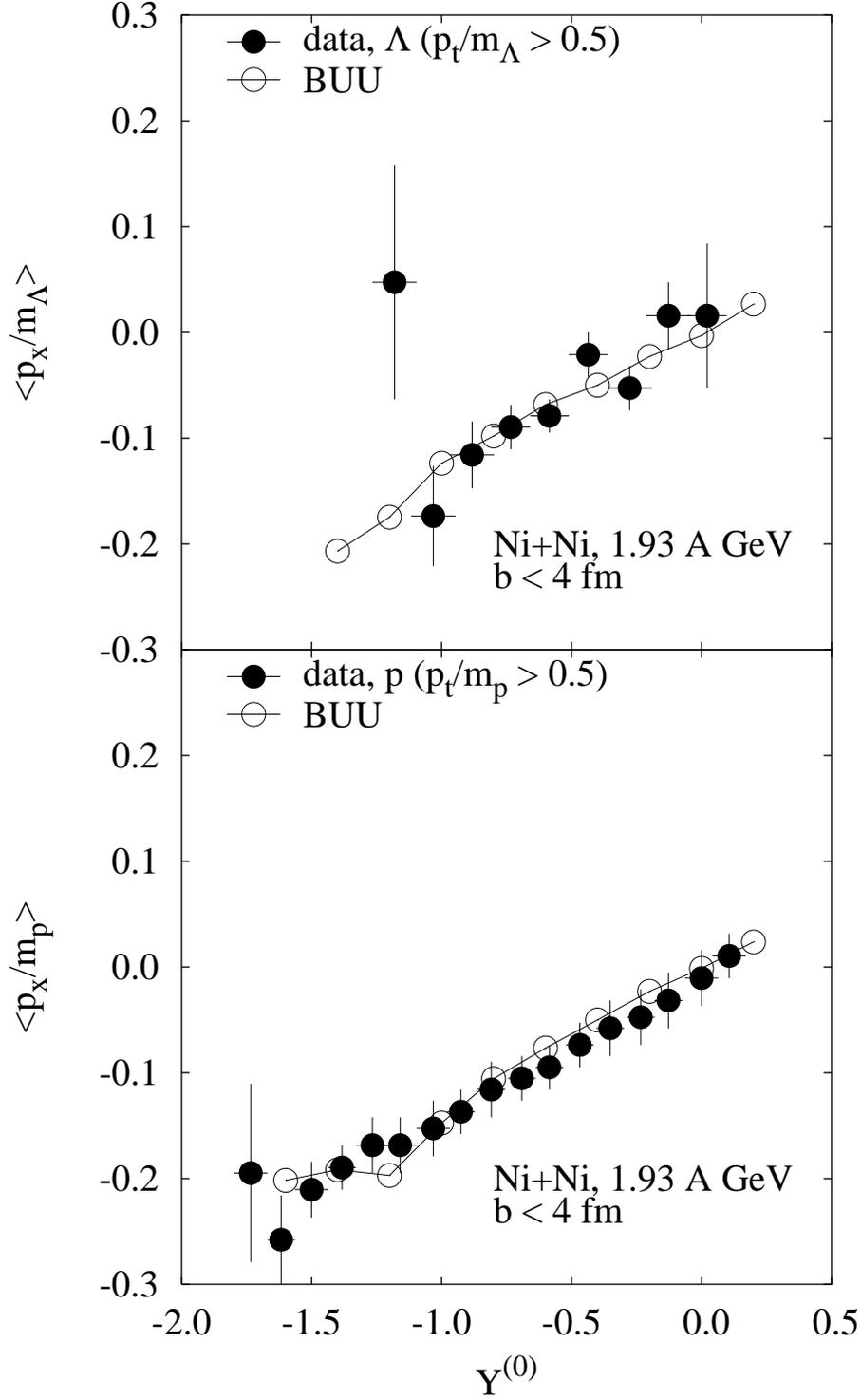}

\caption{\label{fig:pxy_ni193ni_plamsig0_new} Mean transverse momentum 
projected on the reaction plane vs the normalized rapidity for $\Lambda$
hyperons (upper panel) and for protons (lower panel) for   
Ni+Ni collisions at 1.93 A GeV with $b < 4$ fm. The particles are 
selected in the transverse momentum ranges $p_t/m_\Lambda > 0.5$ for
$\Lambda$'s and  $p_t/m_p > 0.5$ for protons. The 
Central Drift Chamber (CDC) angular cuts ($30^o < \Theta_{lab} < 150^o$)
are taken into account in calculations. The data are from Ref. \cite{Rit95}.}

\end{figure}

Fig.~\ref{fig:pxy_ni193ni_plamsig0_new} shows a mean transverse momentum
projected on the reaction plane as a function of a normalized rapidity
$Y^{(0)} \equiv (y/y_{proj})_{c.m.}$ for $\Lambda$ hyperons
(upper panel) and for protons (lower panel) for the system Ni+Ni at
1.93 A GeV. For comparison, we have selected the FOPI data \cite{Rit95}. 
The proton flow ($\equiv d \langle p_x \rangle /dY^{(0)}$ at $Y^{(0)} = 0$) 
is quite well described. In the calculation of the $\Lambda$ flow we have
taken into account both $\Lambda$'s and $\Sigma^0$'s, since in the data 
\cite{Rit95} both kinds of the $\Lambda$ hyperons, those primary and those 
originating from $\Sigma^0 \to \Lambda + \gamma$ decays, are 
indistinguishable. For simplicity, we have neglected, however, a 
difference between the momenta of the decaying  $\Sigma^0$ and 
the outgoing $\Lambda$. A correction for a recoil momentum due
to the photon emission should slightly reduce the calculated flow. 
Nevertheless, we observe a rather good agreement with the data 
on the $\Lambda$-flow.  This supports our choice of the hyperon mean field
(see Sect. II).

It was shown in Ref. \cite{Zheng04} that the FOPI data \cite{Herr99} 
on the $K^+$ in-plane flow can be reasonably well described using the BR
parametrization of the kaon potential. Before discussing an out-of-plane
flow, we will also compare our calculations with the FOPI data 
\cite{Herr99,Crochet00} on the $K^+$ in-plane flow.

Fig.~\ref{fig:pxy_ni193ni_kp} shows the $K^+$ in-plane flow for
Ni+Ni collisions at 1.93 A GeV in comparison to the data from 
Ref. \cite{Herr99}. In full agreement with earlier calculations
of Ref. \cite{Zheng04} we observe that the BR parametrization 
describes the data best (the solid line with open circles).
The KL parametrization (dotted line with open squares) does not give
an enough repulsion to get the negative flow. Neglecting the space component
${\bf V}$ of the kaon vector field in both calculations, i.e. using
static potentials, results in too negative flow for both the
BR (solid line with open pentagons) and the KL (dotted line with open diamonds)
parametrizations. This was also pointed out earlier in \cite{Zheng04}.
The compensation of the repulsive static potential by the ${\bf V}$ field
is so strong, that, e.g. the KL parametrization  gives practically the
same (positive) flow as the calculation without kaon potential (dashed line 
with open triangles) !

An azimuthal distribution with respect to the reaction plane can be
represented by a Fourier expansion~:
\begin{equation}
    {dN \over d\phi}(\phi) 
     \propto 1 + \sum_{n=1}^\infty\, 2v_n\cos(n\phi)~.   
                                                         \label{Fourier}
\end{equation}
The first two coefficients, $v_1$ and $v_2$, in Eq.(\ref{Fourier}) are called 
directed and elliptic flow, respectively, and are expressed as follows~:
\begin{eqnarray}
  v_1 & = & \langle \cos \phi \rangle 
        =   \langle p_x/p_t \rangle                      \label{v1} \\
  v_2 & = & \langle \cos(2\phi) \rangle 
        =   \langle (p_x^2 - p_y^2)/p_t^2 \rangle~.       \label{v2}
\end{eqnarray} 
By neglecting the terms with $n \geq 4$ in (\ref{Fourier}), 
the elliptic flow can be related to a ratio of particle numbers emitted
out ($\phi=\pm 90^o$) and in ($\phi=0^o$ and $180^o$) the reaction plane~:
\begin{equation}
  R = { dN/d\phi(90^0) + dN/d\phi(-90^0) \over
        dN/d\phi(0^0) + dN/d\phi(180^0) } \simeq { 1-2v_2 \over 1+2v_2 }~.
                                                          \label{R}
\end{equation}

\begin{figure}

\includegraphics{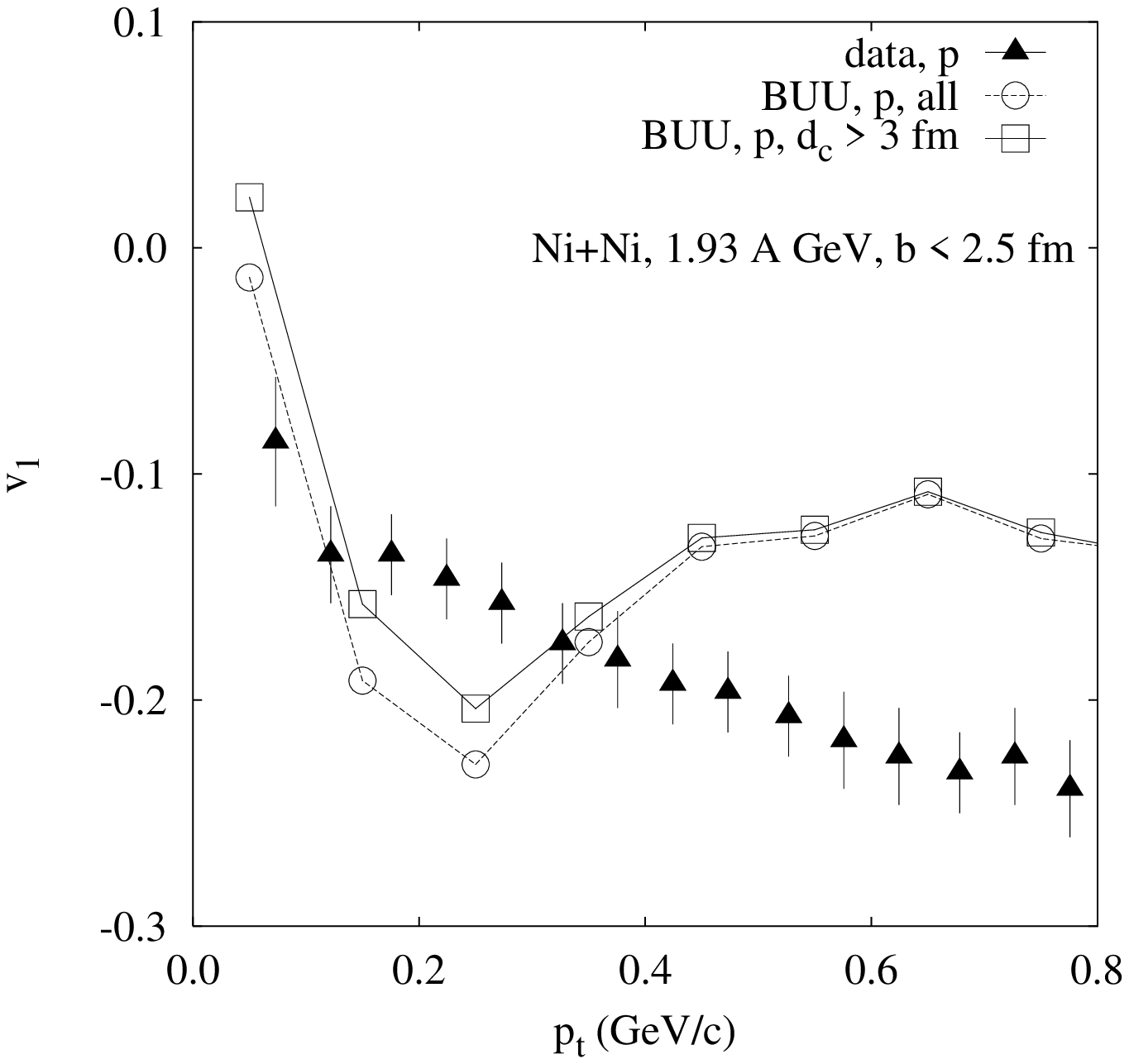}

\caption{\label{fig:v1pt_ni193ni_p} Proton directed flow vs transverse
momentum for collisions Ni+Ni at 1.93 A GeV with $b < 2.5$ fm.
Protons are selected in the rapidity range $-1.2 < Y^{(0)} < -0.65$.
Data are from Ref. \cite{Crochet00}.}

\end{figure}

Fig.~\ref{fig:v1pt_ni193ni_p} shows the directed flow $v_1$ as a function 
of the transverse momentum $p_t$ for protons around the target rapidity from
semicentral Ni+Ni collisions at 1.93 A GeV. We, first, considered
all the protons including those bound in the target spectator remnant 
(dashed line with open circles). Then we selected only the protons 
separated from other particles by a critical distance $d_c > 3$ fm
(solid line with open squares. The proton directed flow is negative 
in agreement with Fig.~\ref{fig:pxy_ni193ni_plamsig0_new}. We observe a 
sensitivity of $v_1$ at low $p_t$'s to the selection of protons: the 
calculation including only the separated protons gives a somewhat smaller 
absolute value of $v_1$ than the calculation including all protons. The 
directed flow at large $p_t$'s is independent of the proton selection 
procedure, and we fail to describe the data here. We ascribe this problem to 
a too hard momentum dependence of the nucleon mean field at large momenta 
\cite{LCGM00}, which pushes the high-momentum protons too early from the 
system, before the directed flow develops. We do not expect, however, that 
this drawback influences kaon propagation, since the high-$p_t$ nucleons are 
not abundant and their contribution to the kaon mean field is small.

Fig.~\ref{fig:v1pt_ni193ni_kp} displays the directed flow of kaons
near target rapidity as a function of $p_t$ for the semicentral collisions 
Ni+Ni at 1.93 A GeV. 
In the upper panel of Fig.~\ref{fig:v1pt_ni193ni_kp} we present 
by solid lines the calculations employing the BR parametrization 
(open circles --- full calculation, open pentagons --- without ${\bf V}$ 
field) and by dotted lines --- the calculations employing
the KL parametrization (open squares --- full calculation, 
open diamonds --- without ${\bf V}$ field). One can see, that the $v_1$ 
coefficient of $K^+$'s is very sensitive to the choice of the kaon potential.
As in the case of $\langle p_x \rangle$ vs $Y^{(0)}$ (c.f. 
Fig.~\ref{fig:pxy_ni193ni_kp}), the BR parametrization provides the best 
description of the data. The KL parametrization gives a too small value 
of $v_1$ indicating not enough repulsion. Neglecting the ${\bf V}$ field 
results in too large values of $v_1$ for the both parametrizations. 

In the lower panel of Fig.~\ref{fig:v1pt_ni193ni_kp} we explore 
the relative importance of the kaon-nucleon (KN) scattering and
of the kaon potential for the description of the directed flow. To this aim,
we have performed three additional calculations: (i) keeping the BR kaon
mean field, but without the KN scattering (solid line with open down 
triangles), (ii) without kaon mean field, but with the KN scattering
(dashed line with open up triangles), and (iii) without kaon mean field
and without the KN scattering (dash-dotted line with stars).
We see that without the kaon mean field it is impossible to reproduce the 
measured positive $v_1$ at small transverse momenta.
The $v_1$ vs $p_t$ dependence for kaons is similar to the one for protons
(c.f. Fig.~\ref{fig:v1pt_ni193ni_p}) in this case.
The kaon mean field alone gives already the correct value of $v_1$.
The KN scattering reduces $v_1$ slightly. This is expected, since the KN 
collisions should make kaons to ``flow'' together with nucleons.

\begin{figure}

\includegraphics{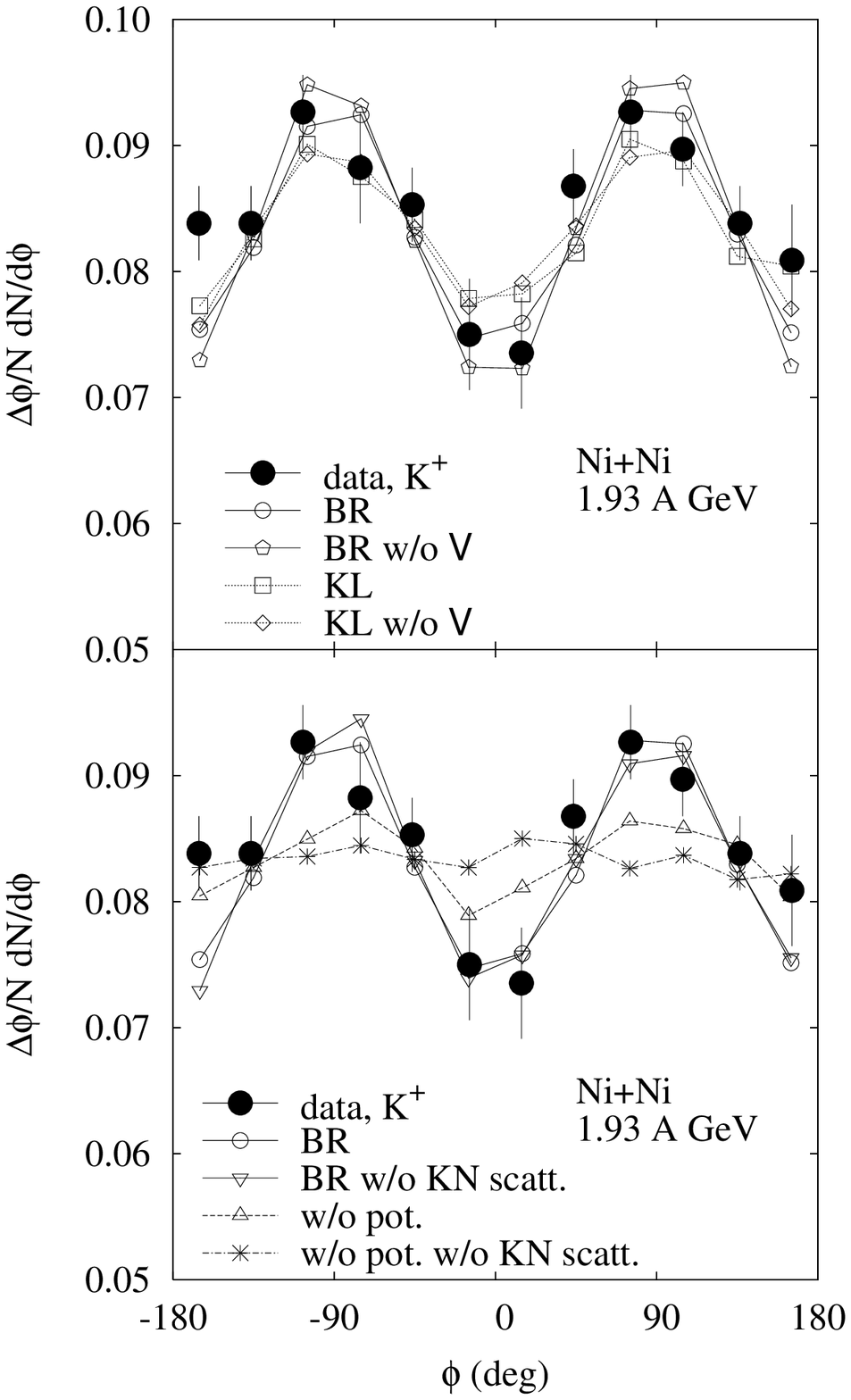}

\caption{\label{fig:azdst_ni193ni} $K^+$ azimuthal distributions for
semicentral ($b=4-6.5$ fm) Ni+Ni collisions at 1.93 A GeV. 
Kaons are selected in the rapidity range $|Y^{(0)}|<0.4$ and in the
transverse momentum range $p_t=0.2-0.8$ GeV/c.
Different calculations are explained in the text.
Data are from Ref. \cite{Uhlig04}.}

\end{figure}

\begin{figure}

\includegraphics{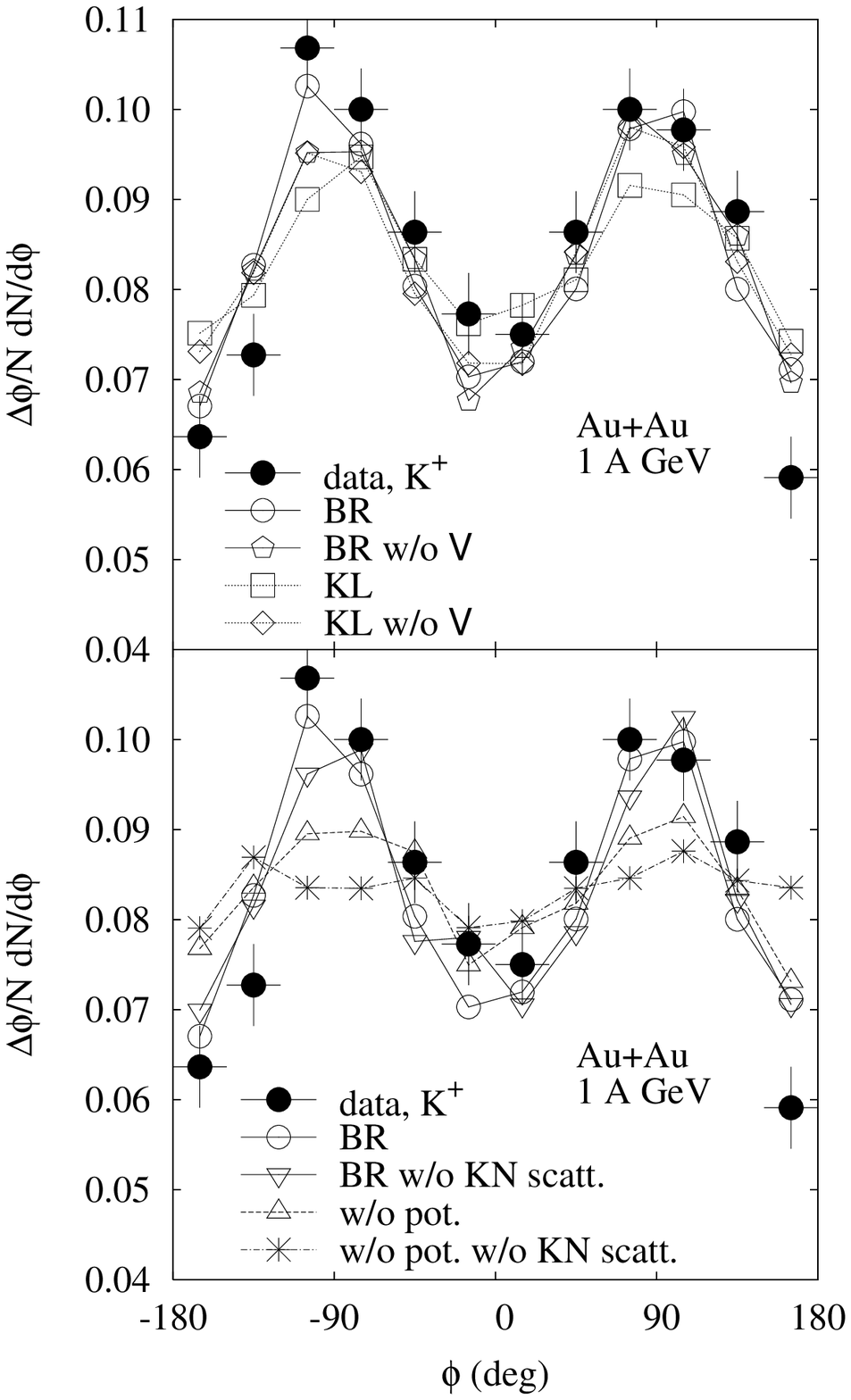}

\caption{\label{fig:azdst_au100au} 
$K^+$ azimuthal distributions for semicentral ($b=5-10$ fm) Au+Au collisions 
at 1 A GeV. Kaons are selected in the rapidity range $|Y^{(0)}|<0.6$ and in 
the transverse momentum range $p_t=0.2-0.8$ GeV/c.
Different calculations are explained in the text.
Data are from Ref. \cite{Shin98}.}

\end{figure}

\begin{figure}

\includegraphics{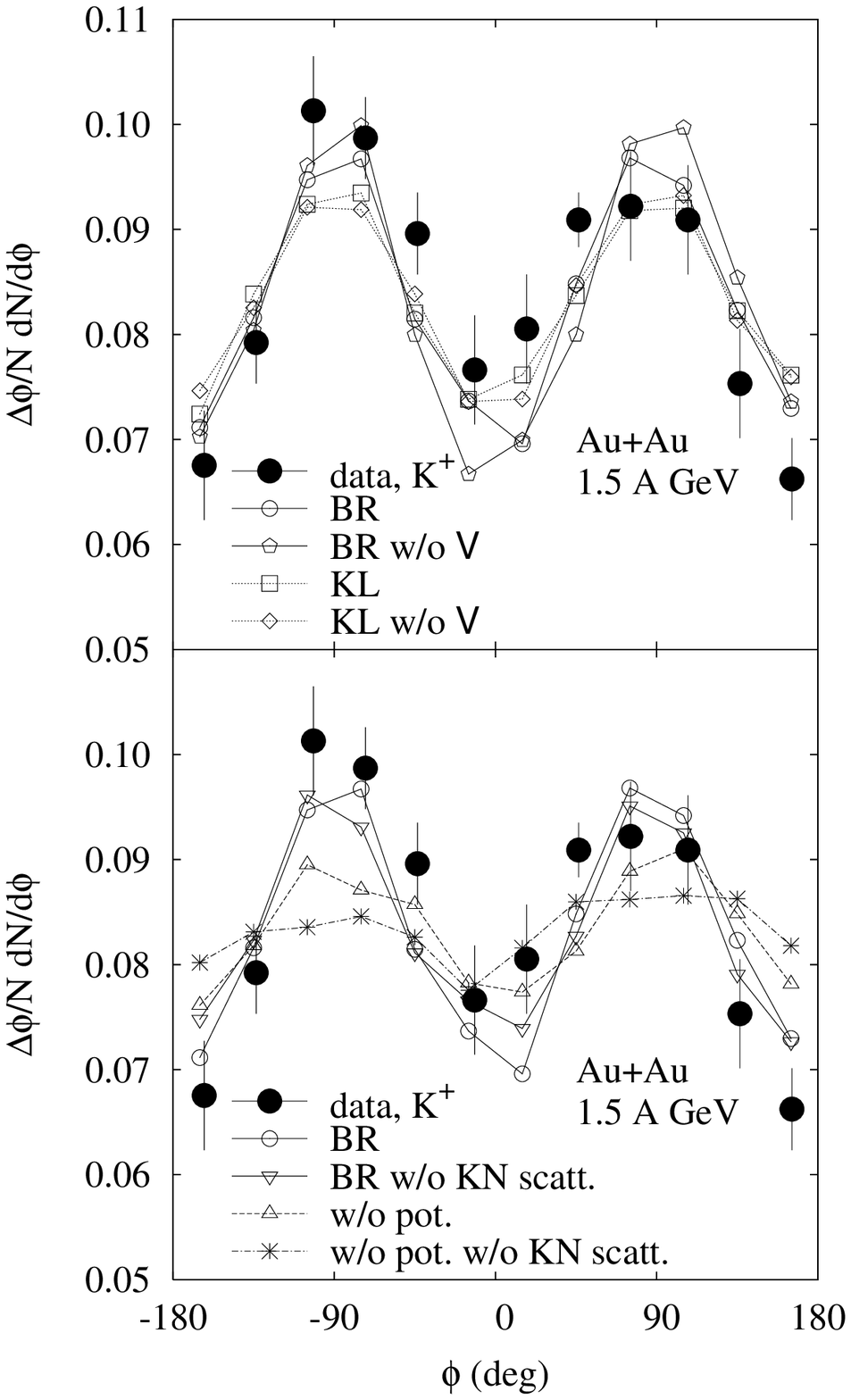}

\caption{\label{fig:azdst_au148au}
$K^+$ azimuthal distributions for semicentral ($b=6-10$ fm) Au+Au collisions 
at 1.5 A GeV. Kaons are selected in the rapidity range $|Y^{(0)}|<0.4$ and in 
the transverse momentum range $p_t=0.2-0.8$ GeV/c.
Different calculations are explained in the text.
Data are from Ref. \cite{Uhlig04}.}

\end{figure}

Figs.~\ref{fig:azdst_ni193ni}, \ref{fig:azdst_au100au} and 
\ref{fig:azdst_au148au} show the azimuthal distributions of $K^+$'s 
at midrapidity from semicentral collisions Ni+Ni at 1.93 A GeV, Au+Au at 
1 A GeV and Au+Au at 1.5 A GeV, respectively, in comparison to the data from 
Refs. \cite{Shin98,Uhlig04}.
In order to quantify an anisotropy of the azimuthal distributions we
have performed a fit of these distributions as in 
Refs. \cite{Shin98,Uhlig04}~:
\begin{equation}
  {dN \over d\phi}(\phi) \propto 1 + 2v_1\cos(\phi) + 2v_2\cos(2\phi)~.
                                                      \label{fit}
\end{equation}
The elliptic flows $v_2$ for the data and for different calculations
are collected in the Table~\ref{tab:v2_kaon} \cite{com3}.

The experimental data for all three systems reveal a pronounced 
out-of-plane ($v_2 < 0$) emission of $K^+$'s. Overall, the calculations
with the BR parametrization of the kaon mean field (solid lines with
open circles) provide the most reasonable description of the data on
the azimuthal distributions. The calculations with the KL parametrization
(dotted lines with open boxes) produce not enough anisotropy.

In order to understand the mechanism of the out-of-plane $K^+$ enhancement
better, we have performed additional calculations by switching-off
various effects. The calculations without the space component of the kaon
vector field ${\bf V}$  (solid lines with open pentagons for the BR
parametrization and dotted lines with open diamonds for the KL one) differ
only very slightly from the full calculations, contrary to the findings
of Ref. \cite{Wang99} on the influence of the Lorentz force on the
kaon squeeze-out. We recall, that for the in-plane flow (c.f. 
Figs.~(\ref{fig:pxy_ni193ni_kp}),(\ref{fig:v1pt_ni193ni_kp})) there is a 
strong influence of the ${\bf V}$ field, in agreement with the
results of Ref. \cite{Zheng04}.

The squeeze-out of particles can be caused either by a dynamical focusing
due to a repulsive mean field \cite{Dani00} or by a shadowing of in-plane
emitted particles by spectator pieces \cite{LCGM00}. Shadowing implies
a dominant role of scattering and/or absorption of the particles 
on nucleons. What determines the kaon squeeze-out: the kaon mean field
or the KN scattering?  To answer this question, as in the case of
the directed flow, we present in the lower panels 
of Figs.~\ref{fig:azdst_ni193ni},\ref{fig:azdst_au100au},%
\ref{fig:azdst_au148au} the results (i) with the BR parametrization, but 
without the KN scattering, (ii) without kaon mean field, but with the KN 
scattering, and (iii) without kaon mean field and without the KN scattering.
(the meaning of the lines is the same as in the lower panel of 
Fig.~\ref{fig:v1pt_ni193ni_kp} discussed above). In the calculation (iii)
there is no squeeze-out signal. The calculations (i) and (ii) both
produce the squeeze-out. However, in the first case the signal is
stronger than in the second case. Futhermore, the calculation (i) gives
almost the same azimuthal distribution and the elliptic flow $v_2$ as
the full calculation including the BR parametrization and the KN scattering.
In the case (i) the mechanism of the out-of-plane enhancement can be only
the dynamical focusing, while in the case (ii) only the shadowing is 
active. Thus, in our calculations the squeeze-out of $K^+$'s is caused
mainly by the mechanism of the dynamical focusing by the repulsive mean 
field.

\begin{figure}

\includegraphics{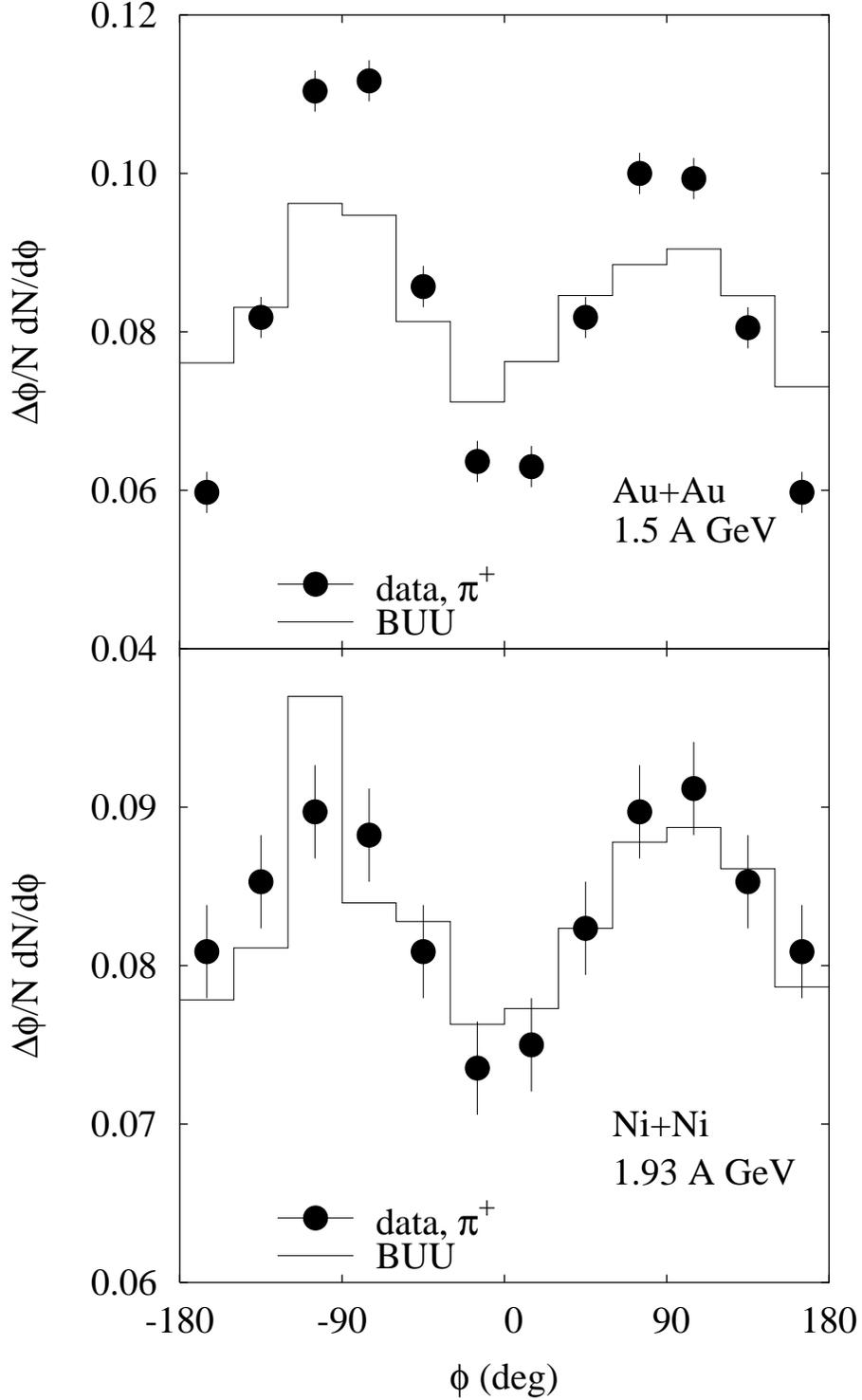}

\caption{\label{fig:azdst_pip} Azimuthal distributions of $\pi^+$-mesons
produced in semicentral ($b=6-10$ fm) Au+Au collisions at 1.5 A GeV 
(upper panel) and semicentral ($b=4-6.5$ fm) Ni+Ni collisions at 1.93 A GeV 
(lower panel). Pions are selected in the rapidity range $|Y^{(0)}|<0.4$ and in 
the transverse momentum range $p_t=0.2-0.8$ GeV/c.
Data are from Ref. \cite{Uhlig04}.}

\end{figure}

For comparison with kaons we, finally, present in Fig.~\ref{fig:azdst_pip}
the azimuthal distributions of the $\pi^+$'s at midrapidity from 
the semicentral collisions Au+Au at 1.5 A GeV and Ni+Ni at 1.93 A GeV.
The corresponding elliptic flows $v_2$ --- obtained in a similar fit
procedure as discussed above in the case of kaons --- are given in 
the Table~\ref{tab:v2_pip}. Since a pion potential is not included in our 
calculations, we can not expect a good agreement with the data 
\cite{Uhlig04}. Indeed, in the case of Au+Au at 1.5 A GeV we  
underestimate the measured squeeze-out signal. The same result was observed 
earlier in Ref. \cite{LCLM01} for the case of Au+Au at 1 A GeV. For the 
lighter system and higher energy (Ni+Ni at 1.93 A GeV) the influence of the 
pion mean field should be smaller and here our calculation is able to reproduce
the data \cite{Uhlig04}.  

\section{ Summary and discussion }

We have studied the $K^+$ production and propagation in heavy-ion collisions
at the SIS energies on the basis of the BUU model \cite{EBM99,MEPhD,LM03}.
The model propagates explicitly a large set of the $N^*$ and $\Delta$
resonances and includes the two main kaon production channels which
are relevant ones at the beam energies of 1-2 A GeV~:
$B B \to B Y K$ and $\pi B \to K Y$. The potentials of a nonstrange baryon
$B$, of a hyperon $Y$ and (optionally) of a kaon $K$ were included in
the cross sections via shifts of the thresholds. For the kaon potential
we have used the BR and the KL parametrizations, which have been already
applied in Ref. \cite{Zheng04}. The KL parametrization
can be derived from the lowest-order approximation to the chiral
Lagrangian \cite{Li951}. The BR parametrization \cite{BR96}, in distinction
to the KL one, takes into account the in-medium pion decay constant in
the kaon vector field and also has a larger kaon-nucleon sigma term.

\begin{figure}

\includegraphics{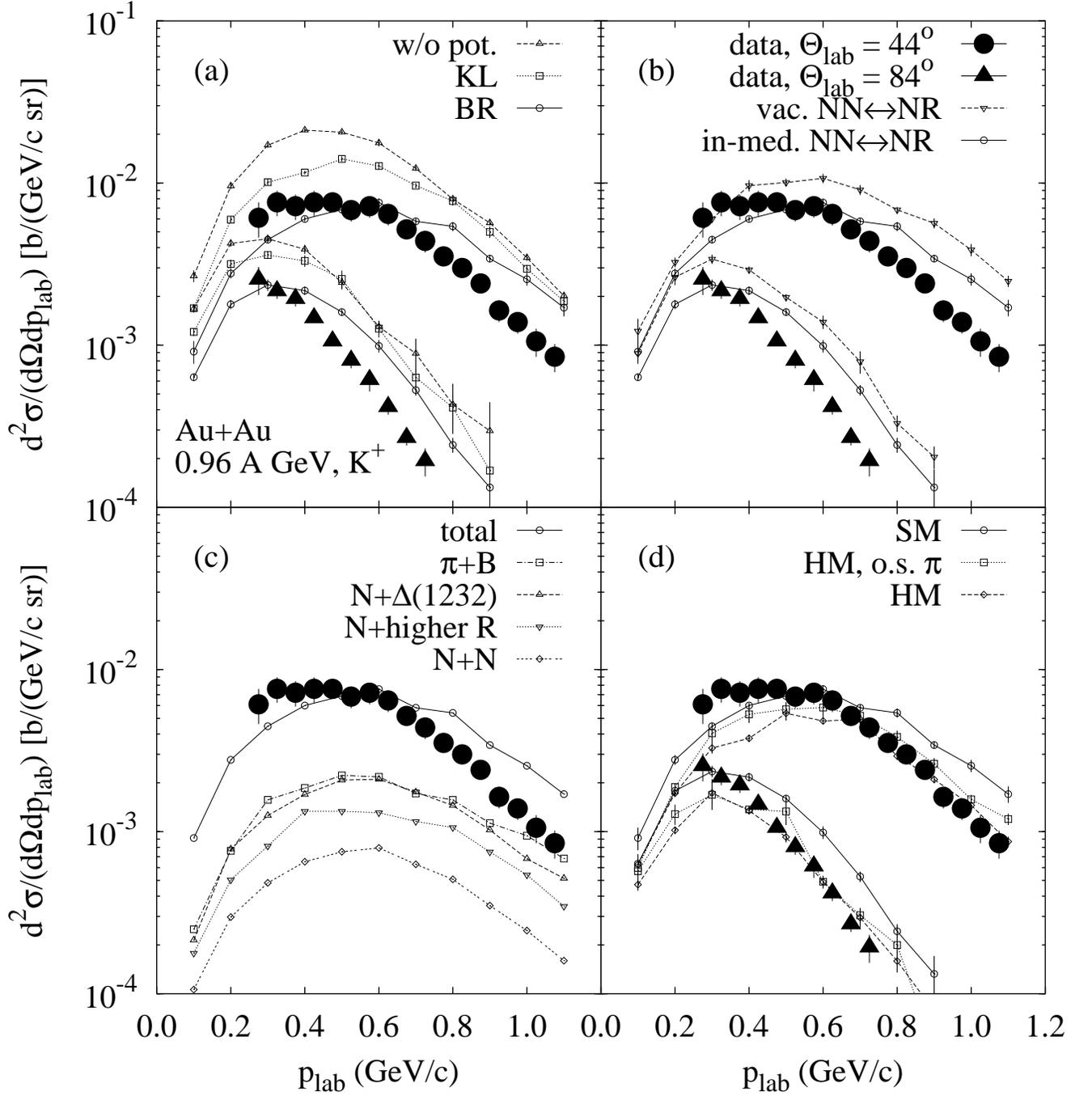}

\caption{\label{fig:d2sig_au096au} Inclusive K$^+$ production
cross section as a function of the laboratory momentum for Au+Au
collisions at 0.96 A GeV at various laboratory angles. Errorbars on
the calculated results are statistical. See text for further 
details of calculations. Data are from Ref. \cite{Sturm01}.}

\end{figure}

\begin{figure}

\includegraphics{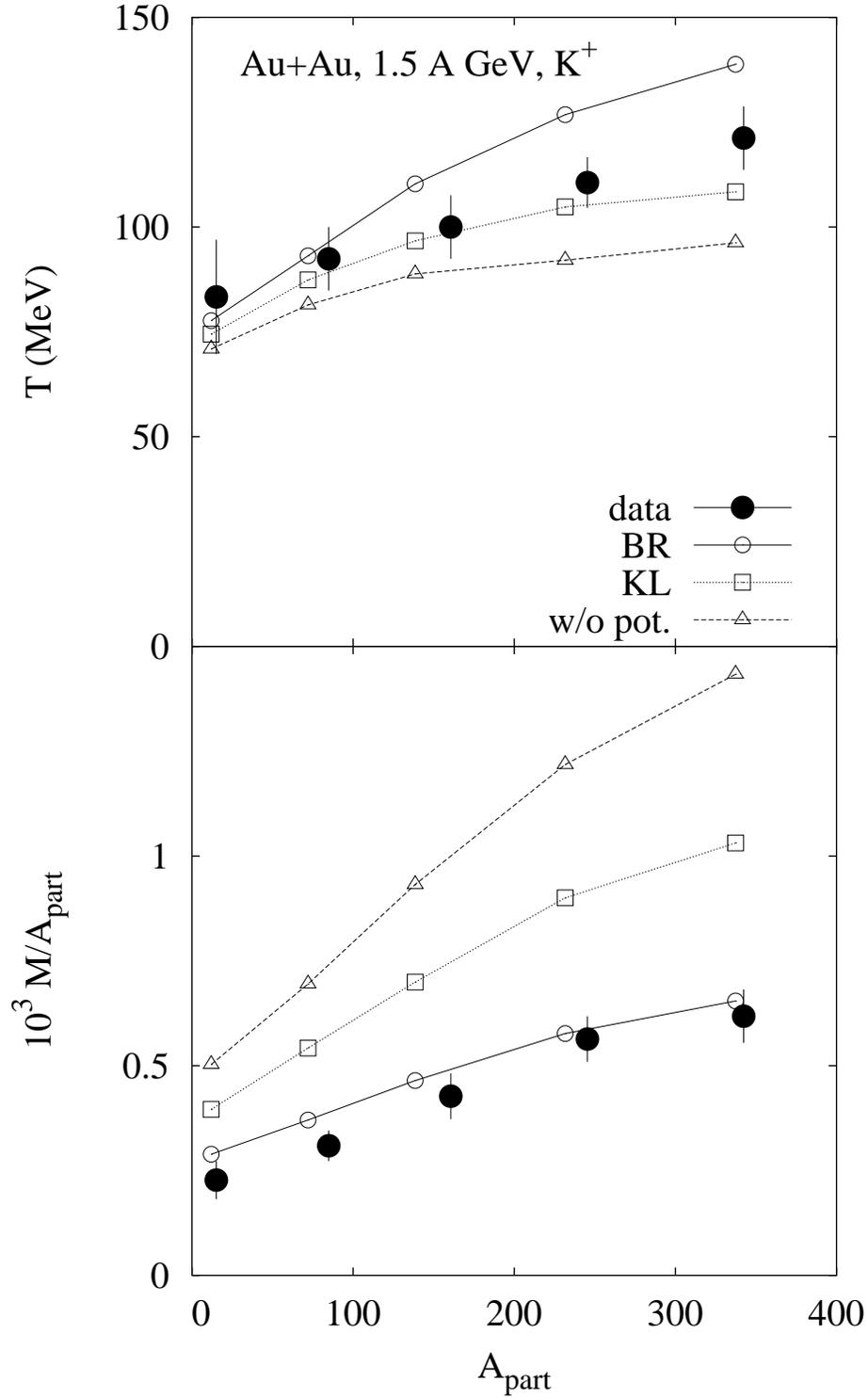}

\caption{\label{fig:slope_au150au_kp} Upper panel: inverse slope parameter 
of the K$^+$ c.m. kinetic energy spectrum as a function of the number of 
participating nucleons $A_{part}$ for the system Au+Au at 1.5 A GeV. 
Lower panel: K$^+$ multiplicity per participating nucleon vs $A_{part}$
for Au+Au at 1.5 A GeV. The data are from Ref. \cite{Forst03}.}

\end{figure}

First, we have looked at the $K^+$ yields and spectra. The secondary 
processes $R N \to B Y K$ and $\pi B \to K Y$ have a big contribution
to the kaon production (see the Table~\ref{tab:channels} and
Fig.~\ref{fig:d2sig_au096au}c). This makes the kaon yield very sensitive
to the in-medium reduction of the $N N \leftrightarrow N R$ cross sections.
In our ``standard'' calculation, which includes the SM nucleon mean field,
the in-medium $N N \leftrightarrow N R$ cross sections \cite{LM03} and
the kaon mean field in the BR parametrization, we have obtained a very 
good description of the kaon multiplicities for both heavy and light
colliding systems in all the SIS energy region, in agreement with
the QMD calculations of Ref. \cite{Fuchs01}.
For a light system C+C at 2 A GeV the $p_{lab}$-spectra are reproduced quite 
well, which is also shown in Ref. \cite{Zheng04}. However, for heavy systems
Au+Au and Ni+Ni the calculated $p_{lab}$-, $E_{kin}^{cm}$- and 
$m_T$-spectra are somewhat too hard. The repulsive kaon potential
reduces the yield of soft kaons and leaves the yield of hard kaons
practically unchanged. Thus, the kaon potential makes the kaon spectra harder.
In most cases the slopes of the $K^+$ spectra are reasonably well described
by the calculation without kaon potential. But this calculation drastically
overestimates the kaon multiplicity. Moreover, the slopes of the 
$E_{kin}^{cm}$-spectra for the central collisions of the heaviest system 
Au+Au at 1.5 A GeV can be {\it only} described by using the repulsive kaon 
potential (Fig.~\ref{fig:slope_au150au_kp}).

Second, we have studied the collective in-plane and out-of-plane flows
of $K^+$'s. The $\langle p_x \rangle$ vs $Y^{(0)}$ and $v_1$ vs $p_t$ 
dependencies for kaons emitted from semicentral Ni+Ni collisions at 1.93 A GeV
are very sensitive to the choice of the kaon mean field and clearly
favour the BR parametrization, as was also recently shown in 
Ref. \cite{Zheng04}. The azimuthal distributions of kaons at midrapidity
are also sensitive to the choice of the kaon mean field. The KN scattering
alone gives a too weak squeeze-out signal with respect to the data.
For the heaviest measured system Au+Au at 1 and 1.5 A GeV, the data on the 
elliptic flow $v_2$ are best described by using the BR parametrization.
For a lighter system Ni+Ni at 1.93 A GeV the calculation with the KL 
parametrization reproduces the data on $v_2$ best, but the
BR one is also consistent with the data within the experimental errorbars.

\begin{figure}

\includegraphics{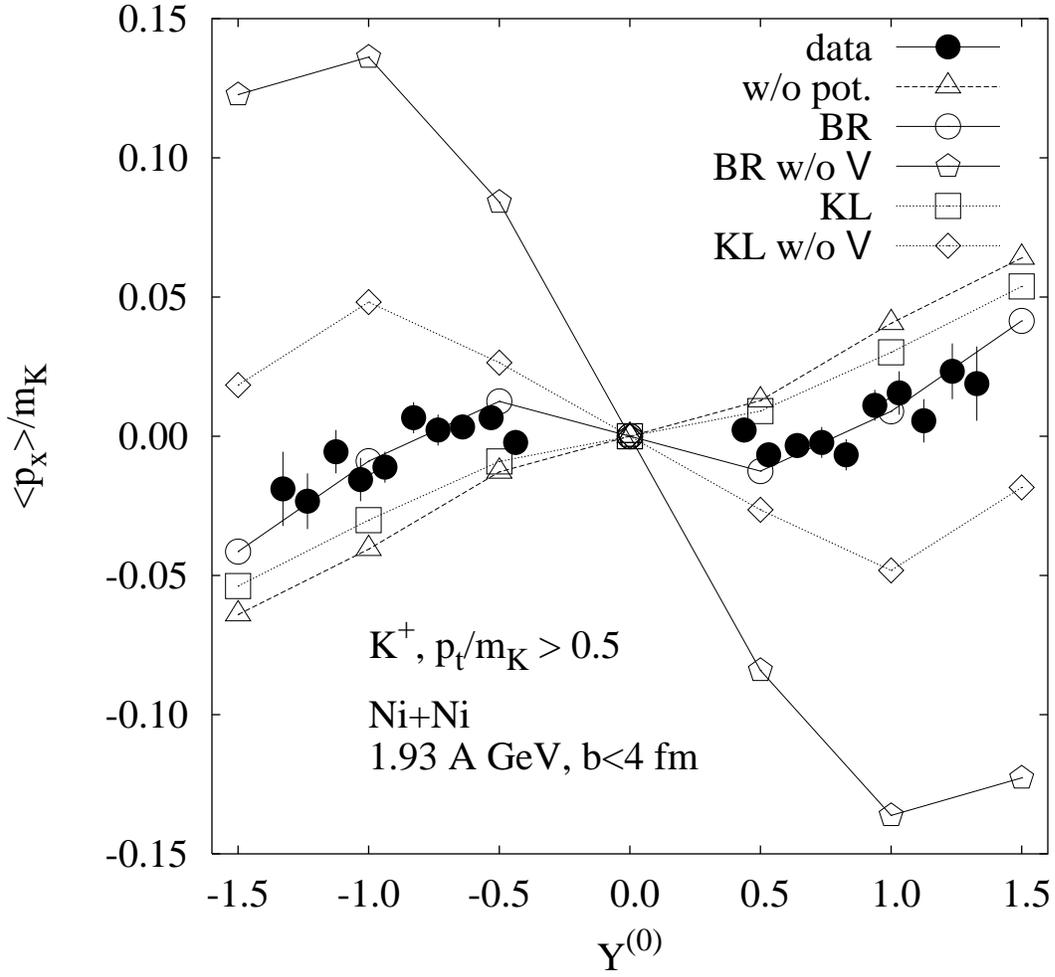}

\caption{\label{fig:pxy_ni193ni_kp} $K^+$ mean transverse momentum projected
on the reaction plane vs the normalized rapidity for Ni+Ni at 1.93 A GeV,
$b < 4$ fm compared with the data from Ref. \cite{Herr99}. 
Only high transverse momentum kaons ($p_t/m_K > 0.5$) were selected according 
to \cite{Herr99}. The CDC angular cuts 
($39^o < \Theta_{lab} < 150^o$) and the upper limit of the laboratory
momentum (0.5 GeV/c) up to which a $K^+$ can be identified \cite{Herr99}
were also taken into account in calculations. 
The data points and the calculated curves at $Y^{(0)} > 0$ 
are obtained by reflection from those at $Y^{(0)} < 0$.}

\end{figure}

\begin{figure}

\includegraphics{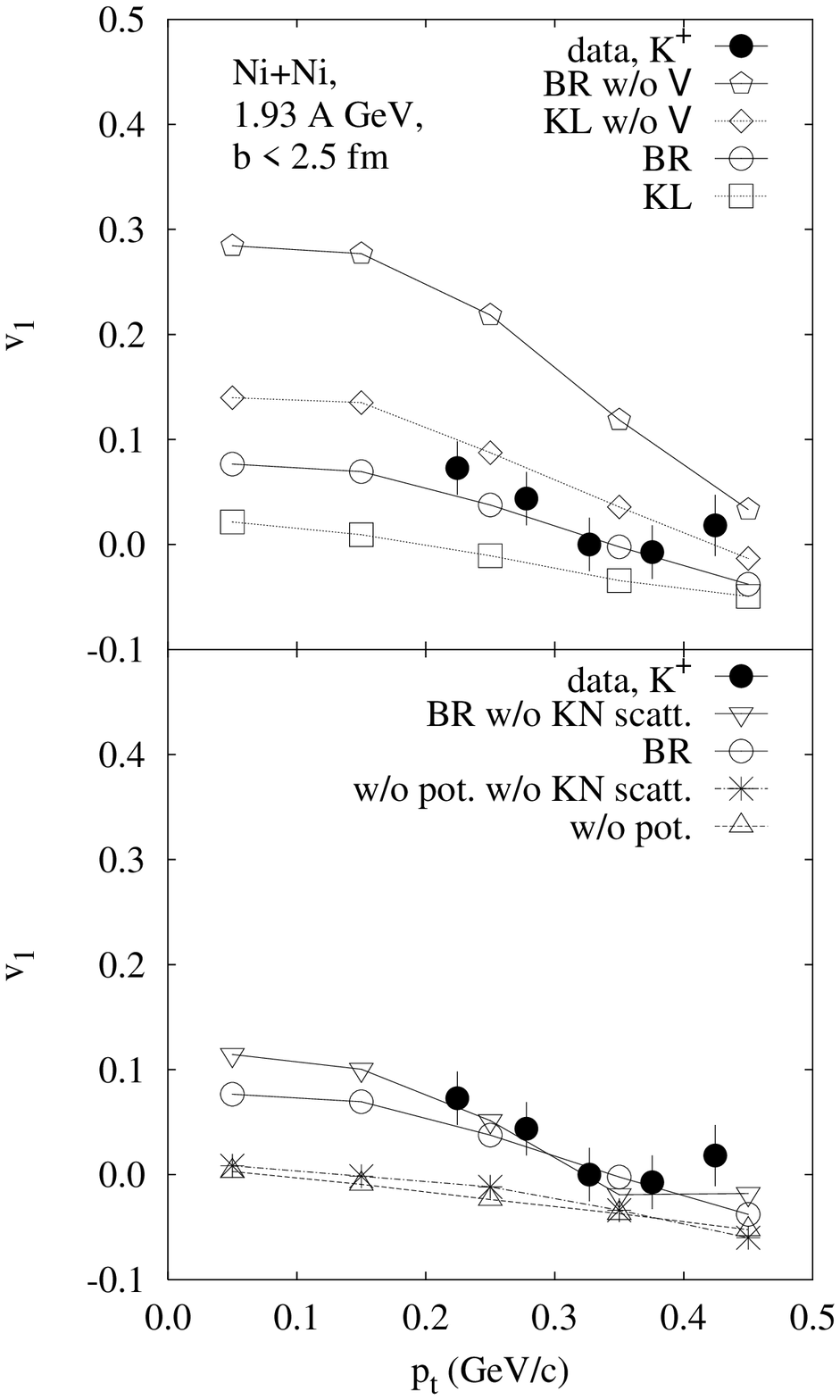}

\caption{\label{fig:v1pt_ni193ni_kp} $K^+$ directed flow vs transverse
momentum for collisions Ni+Ni at 1.93 A GeV with $b < 2.5$ fm.
Kaons are selected in the rapidity range $-1.2 < Y^{(0)} < -0.65$.
Data are from Ref. \cite{Crochet00}.}

\end{figure}

The Lorentz force caused by the space components ${\bf V}$ of the kaon
vector field does not significantly influence the azimuthal distributions
of kaons at the midrapidity. However, in the case of the in-plane flow
the Lorentz force is found to contribute very strongly \cite{Zheng04}
(see also our Figs.~\ref{fig:pxy_ni193ni_kp},\ref{fig:v1pt_ni193ni_kp}).
Thus, the kaon azimuthal distributions at midrapidity probe, basically,
the static kaon potential.

Some comments are also in order with regard to our results of a recent
benchmark test of the transport codes \cite{THW04}. Our results in
\cite{THW04} were obtained with an enforced $\Delta(1232)$ lifetime
of $\hbar/120$ MeV and with the the vacuum $NN \leftrightarrow NR$
cross sections. In our standard calculations, however, we use the
$\Delta(1232)$ lifetime of $\hbar/\Gamma_\Delta(M)$, where 
$\Gamma_\Delta(M)$ is the mass dependent width of the $\Delta(1232)$-resonance
\cite{EBM99}, and the in-medium $NN \leftrightarrow NR$ cross sections.
In particular, using the vacuum $NN \leftrightarrow NR$ cross sections
in \cite{THW04} has led to enhanced pion and kaon yields in our
calculations with respect to most other transport calculations.
However, our results \cite{THW04} on $\pi^\pm$'s and $K^+$'s agree with those 
obtained with the BUU code of Refs. \cite{Wolf93,Wolf97,Barz03}, which also 
propagates explicitly a large set of the baryonic resonances with $M < 2$ GeV.
Thus, our present study demonstrates clearly, that the effect of the higher
baryonic resonances on the pion and kaon production is counterbalanced
by the in-medium reduction of the $NN \leftrightarrow NR$ cross sections,
which produces a satisfactory agreement with experimental data.

In conclusion, the BR parametrization of the kaon mean field provides
the best overall description of the $K^+$ observables at SIS energies.
The remaining problem lies in the somewhat too hard $p_{lab}$-spectra for 
the Au+Au collisions. These spectra, however, are sensitive not only
to the choice of the kaon mean field, but also to the kaon production
cross sections in nuclear medium, which are still rather ambiguous.
The in-medium calculation for the channel $\pi B \to Y K$ has been performed 
in Ref. \cite{Tsush00} indicating a reduction of the cross section 
at a finite baryon density. Such an in-medium calculation is still needed, 
however, for another important channel $B B \to B Y K$, basing e.g. on the 
model of Ref. \cite{Tsush99}.

\begin{acknowledgments}
We gratefully acknowledge support by the Frankfurt Center for Scientific 
Computing.
\end{acknowledgments}

\newpage

\begin{table*}
\caption{\label{tab:channels} Contribution of the different kaon 
production channels to the total kaon number from central ($b=0$ fm) 
collisions of the various systems.}
\begin{ruledtabular}
\begin{tabular}{lccc}
 Channel        & Au+Au, 0.96 A GeV  & Au+Au, 1.48 A GeV & Ni+Ni, 1.93 A GeV\\
\hline
$R N \to B Y K$ &    51\%            &    47\%           &   36\%          \\
$N N \to B Y K$ &    12\%            &    25\%           &   47\%          \\
$\pi B \to K Y$ &    25\%            &    17\%           &   10\%          \\
$R R \to B Y K$ &    8\%             &    6\%            &   3\%           \\
$\pi B \to K \bar K N$ & 3\%         &    3\%            &   2\%           \\ 
$B B \to N N K \bar K$ & 1\%         &    2\%            &   2\%           \\
\end{tabular}
\end{ruledtabular}
\end{table*} 
 
\newpage

\begin{table*}
\caption{\label{tab:v2_kaon} The elliptic flow $v_2$ obtained by the fit
of the $K^+$ azimuthal distributions by a function 
$\propto 1 + 2v_1\cos(\phi) + 2v_2cos(2\phi)$. 
The data are from Refs. \cite{Shin98,Uhlig04}.} 
\begin{ruledtabular}
\begin{tabular}{llll}
                & Au+Au, 1 A GeV   & Au+Au, 1.5 A GeV & Ni+Ni, 1.93 A GeV \\
\hline
Exp.            & -0.110 $\pm$ 0.011 & -0.08 $\pm$ 0.02   & -0.04 $\pm$ 0.02 \\
BR              & -0.100 $\pm$ 0.007 & -0.082 $\pm$ 0.004 & -0.059 $\pm$ 0.002 \\
BR w/o {\bf V}  & -0.092 $\pm$ 0.005 & -0.098 $\pm$ 0.006 & -0.076 $\pm$ 0.001 \\
BR w/o KN scatt.& -0.088 $\pm$ 0.011 & -0.068 $\pm$ 0.006 & -0.061 $\pm$ 0.003 \\
KL              & -0.055 $\pm$ 0.005 & -0.062 $\pm$ 0.003 & -0.037 $\pm$ 0.004 \\
KL w/o {\bf V}  & -0.080 $\pm$ 0.005 & -0.062 $\pm$ 0.003 & -0.041 $\pm$ 0.002 \\
w/o pot.        & -0.048 $\pm$ 0.005 & -0.040 $\pm$ 0.004 & -0.020 $\pm$ 0.002 \\
w/o pot. w/o KN scatt.& -0.015 $\pm$ 0.005 & -0.017 $\pm$ 0.005 & -0.002 $\pm$ 0.002\\
\end{tabular}
\end{ruledtabular}
\end{table*}
 
\begin{table*}
\caption{\label{tab:v2_pip} The elliptic flow $v_2$ obtained by the fit
of the $\pi^+$ azimuthal distributions by a function 
$\propto 1 + 2v_1\cos(\phi) + 2v_2cos(2\phi)$. 
The data are from Ref. \cite{Uhlig04}.} 
\begin{ruledtabular}
\begin{tabular}{lll}
                & Au+Au, 1.5 A GeV  & Ni+Ni, 1.93 A GeV \\
\hline
Exp.            & -0.15  $\pm$ 0.01   & -0.04  $\pm$ 0.01  \\
BUU             & -0.063 $\pm$ 0.006  & -0.041 $\pm$ 0.008 \\
\end{tabular}
\end{ruledtabular}
\end{table*}

\clearpage

\thispagestyle{empty}

\clearpage

\thispagestyle{empty}

\clearpage

\thispagestyle{empty}

\clearpage

\thispagestyle{empty}

\clearpage

\thispagestyle{empty}

\clearpage

\thispagestyle{empty}

\clearpage

\thispagestyle{empty}

\clearpage

\thispagestyle{empty}

\clearpage

\thispagestyle{empty}

\clearpage

\thispagestyle{empty}

\clearpage

\thispagestyle{empty}

\clearpage

\thispagestyle{empty}

\clearpage

\thispagestyle{empty}

\clearpage

\thispagestyle{empty}

\clearpage

\thispagestyle{empty}

\clearpage

\thispagestyle{empty}

\clearpage

\thispagestyle{empty}

\begin{thebibliography}{99}
\bibitem{com1} Please, do not mix these masses with the Dirac masses !
\bibitem{com4} Notice, that in Fig.~\ref{fig:rateint}a and in 
Table~\ref{tab:channels} we have summed up both channels nucleon-$\Delta(1232)$
and nucleon-higher resonance, while in Fig.~\ref{fig:d2sig_au096au}c these 
channels are shown separately.
\bibitem{com3} The parameters $v_1$ and $v_2$ of the fit (\ref{fit}) are 
equal to those given by Eqs.(\ref{v1}),(\ref{v2}) within accuracy up to the 
terms with $n > 2$ in the Fourier expansion (\ref{Fourier}), which we neglect.
Thus, we will also call the fit parameters $v_1$ and $v_2$ as the directed 
and elliptic flows, respectively. Since the azimuthal distributions are 
extracted at midrapidity, the directed flow $v_1$ should be equal to zero 
due to symmetry reasons. We have obtained $v_1 \sim 10^{-3}$ in the fitting 
procedure due to a finite statistics, which is much smaller than the extracted
values of $|v_2|$. Thus, we drop the explicit values of the $v_1$ coefficient.
\end{thebibliography}
\end{document}